# Chemistry of the Earth's Earliest Atmosphere

Bruce Fegley, Jr.[1] and Laura K. Schaefer[1,2]

[1]Planetary Chemistry Laboratory, Department of Earth & Planetary Sciences and

McDonnell Center for the Space Sciences, Washington University, St. Louis, MO 63130

USA, bfegley@wustl.edu

[2]Current address: Department of Astronomy, Harvard University, 60 Garden Street,

Cambridge, MA 02138 USA, lschaefer@cfa.harvard.edu

## Introduction and Overview

In this chapter we describe chemistry of Earth's early atmosphere during and
shortly after its formation where there is little if any geological record. Nevertheless,
current thinking about the silicate vapor, steam, and gaseous stages of atmospheric
evolution on the early Earth is potentially testable by spectroscopic observations of
transiting rocky exoplanets. This assertion is supported by the rapid growth of extrasolar
planetary astronomy from one extrasolar planet in 1995 to 838 as of the time of writing
(Sept. 2012). Over 285 transiting extrasolar planets have been discovered, and
spectroscopic observations have been done for 8 of them and tentatively show one or more of the following gases in their atmospheres – H, Na, K, CO, $CO_2$, $CH_4$, $H_2$, $H_2O$, TiO, and VO (see the catalog on exoplanet.eu). Several groups are actively modeling the chemical composition (e.g., Schaefer et al. 2012) and expected spectral signatures (e.g., Marley et al. 2011) of transiting rocky exoplanets because the rapidly expanding spectroscopic capabilities will allow observational tests in the next few years. Thus, ideas about the Earth's early atmosphere, which cannot be constrained by biological or geological evidence, may be indirectly constrained in the near future by astronomical observations. We return to this point at the end of the chapter.

The nature of Earth's early atmosphere is of particular interest because it is the environment in which life originated sometime between ~ 4.5 and ~ 3.85 Ga ago (Bada 2004). The $^{182}Hf$ – $^{182}W$ and $^{146}Sm$ – $^{142}Nd$ chronometers (e.g., Yin et al 2002, Boyet and Carlson 2005, Moynier et al. 2010) show Earth's core formed ~ 30 million years, and crust formed ~ 38-75 million years, respectively, after formation of the solar system at 4,568 Ma ago (Bouvier and Wadhwa 2010). In other words, the Earth formed ~ 4.54 – 4.49 Ga ago. Norman et al. (2003) derive a crystallization age of 4,456 Ma ago for the earliest lunar crust, i.e., 112 million years after solar system formation. Lunar crust must have formed after the Moon itself formed (by impact of a Mars-sized body with the early Earth). The oldest samples of continental crust are detrital zircons from Jack Hills, Australia, dating back ~ 4,404 Ma ago (Wilde et al. 2001, Harrison et al. 2005). The metavolcanic and metasedimentary rocks from the Nuvvuagittuq (Porpoise Cove) greenstone belt near Hudson Bay in northern Quebec have a $^{146}Sm$ – $^{142}Nd$ isochron age of ~ 4,280 Ma (O'Neil et al. 2009), which probably indicates that these rocks were derived from material extracted from Earth's mantle at that time (O'Neil et al 2009, Sleep 2010, Arndt and Nisbet 2012). The oldest dated rock is the Acasta Gneiss in the Northwest Territories of Canada with an age of 4,031 Ma (Bowring and Williams 1999).

Mojzsis et al. (2001) report oxygen isotopic evidence (from the Jack Hills zircons) for liquid water on Earth's surface ~ 4,300 Ma ago. Another oxygen isotopic study of Jack Hills zircons by Cavosie et al. (2005) suggests liquid water existed on Earth's surface by ~ 4,200 Ma ago and possibly as early as 4,325 Ma ago. This work and other evidence led Arndt and Nisbet (2012) to conclude the Hadean Earth was more clement than previously believed.

Non-unique carbon isotopic signatures possibly indicate photoautotrophic carbon fixation in the Akilia Island banded iron formation (~ 3,850 million years old) in Greenland (Mojzsis et al. 1996, Eiler et al. 1997). Taking the results above at face value the conclusion is that the Earth formed, differentiated, was struck by a giant impactor to form the Moon, and had cooled sufficiently to form continental crust within 164 million years of the formation of the solar system. Whether or not formation of the Earth and/or its differentiation was contemporaneous with lunar formation is unclear. Within another ~ 100-200 Ma liquid water was present on Earth's surface. Finally, sometime in the next ~ 554 million years after continental crust formation primitive life forms originated and evolved to the point where photoautotrophic activity had imprinted its carbon isotopic signature on the early Earth. The carbon isotopic signatures of photoautotrophic carbon fixation are controversial, but at present available evidence indicates the development of life on Earth by ~ 3.8 Ga ago (e.g., see Arndt and Nisbet 2012, Sleep et al. 2012).

As mentioned earlier, we deal with chemistry of the early terrestrial atmosphere during and shortly after formation of the Earth, i.e. within the period between 4,540 and 3,850 Ma. If, as widely believed, the Moon formed by a giant impact on the Earth (Cameron and Ward 1976, Hartmann and Davis 1975), it is likely that any pre-existing atmosphere was destroyed in the aftermath when the Earth was heated to temperatures in excess of 6000 K at the surface and 15,000 K in its interior (Canup 2008). However Abe (2011) and Genda and Abe (2003, 2005) find that most of the oceans and some of the atmosphere should survive a giant impact. If they are correct then outgassing during accretion prior to the Moon-forming impact still plays an important role in Earth's volatile inventory and early atmosphere. This could be very important because of the high proportion (up to 70%) of reduced material (like enstatite chondrites) that is postulated to form the Earth (see our discussion of the sources of volatiles accreted by the Earth).

The plausible scenario which we adopt in this chapter is that the Earth's first atmosphere contained silicate vapor. Water vapor and its thermal dissociation products ($H_2$, $O_2$), volatile gases formed by C (CO, $CO_2$), N ($N_2$, NO), S ($SO_2$, SO, $H_2S$) and other volatile but less abundant species (e.g., KOH, KCl, KF, NaOH, NaCl, NaF, HCl, HF; see Schaefer et al. 2012) partitioned between the silicate vapor atmosphere and a silicate magma ocean on Earth's surface. With further cooling the silicate vapor condensed into molten silicate rain and solid silicate snow and precipitated out of the atmosphere leaving steam, $H_2$, C (CO, $CO_2$), N ($N_2$, NO), S ($SO_2$, $H_2S$) bearing gases, and some other volatile species (HCl, HF, NaCl, KCl) behind. Finally the Earth cooled enough that halogen-bearing minerals and most of the water vapor condensed out of its atmosphere leaving behind only trace water vapor, $H_2$, and C (CO, $CO_2$, $CH_4$), N ($N_2$, $NH_3$) and S ($SO_2$, $H_2S$) bearing gases. This overall picture has been discussed by Sleep and Zahnle in several papers (Sleep and Zahnle 2001, Sleep et al 2001, Sleep 2010, Zahnle 2006, Zahnle et al. 2007, 2010) and modeled by Schaefer et al. (2012).

However, the nature of the $H_2$, C-, N-, S-bearing gas atmosphere, i.e., reducing, neutral, or oxidizing, is controversial, to say the least. In the 1950s experiments by Stanley Miller and Harold Urey produced amino acids, carboxylic acids, HCN, and other organic compounds from a spark discharge in a mixture of $CH_4$, $NH_3$, $H_2$, and $H_2O$ (the latter from the vapor pressure over liquid water) (Miller 1955, Miller and Urey 1959). Miller and Urey chose this reducing atmosphere because spectroscopic observations of Jupiter and Saturn showed $NH_3$ and $CH_4$, and large amounts of $H_2$ were also inferred to be present. The atmospheres of the gas giant planets were interpreted as captured remnants of reducing solar nebula gas. In Urey's model of the origin of the solar system, the primordial atmospheres of Earth, Venus, and Mars were predicted to have been similar (Urey 1952). Furthermore at the time Oparin (1938) and other scientists interested in the origin of life (e.g., Bernal 1949) had also presented arguments in favor of a reducing atmosphere on the early Earth.

In the decades after Miller (1955) published his work, many experiments were done to produce organic compounds from reducing gas mixtures using electrical discharges, heat, or UV light (e.g., see the reviews by Oró et al. 1990, Bada 2004). Taken together these experiments show that Miller-Urey type reactions are most efficient (molecules/J energy input) in reducing $H_2 + CH_4$-bearing atmospheres, are less efficient but still viable in mildly reducing $H_2 + CO$ atmospheres, but significantly less efficient in neutral or oxidizing $H_2O + CO_2$ atmospheres (Oró et al. 1990).

However the reactivity that makes $CH_4$- and $NH_3$-bearing atmospheres so fruitful for organic compound synthesis also makes these gases (especially $NH_3$) subject to rapid photolytic destruction by UV sunlight (e.g., see Abelson 1966, Kuhn and Atreya 1979, Kasting et al. 1983, Zahnle 1986, Tian et al. 2011). Ammonia is destroyed more rapidly than $CH_4$ because $NH_3$ absorbs UV photons shortward of 235 nm in the region of the solar spectrum where the UV flux is much larger than at shorter wavelengths. Methane is photolyzed only by UV photons shortward of 160 nm where the Lyman $\alpha$ line at 121.6 nm provides most of the flux.

Atmospheric chemistry on Titan, Saturn's largest satellite, led different groups to propose a solution for the photochemical instability of a reducing atmosphere on the early Earth. Titan's surface is shrouded from view by high altitude photochemically generated hazes. These are produced by solar UV photochemical destruction of $CH_4$ in Titan's atmosphere. Yet Titan's $N_2$-rich atmosphere is reducing with 1-5 % $CH_4$, and ~ 960 ppmv $H_2$. Sagan and Chyba (1997) proposed that hydrocarbon haze on the early Earth protected $NH_3$ from complete destruction by UV sunlight, leading to a steady state mole fraction of $10^{-5\pm1}$, which also provides greenhouse warming to counteract the faint early Sun. Although their results were later contradicted by Pavlov et al (2001), the idea of a hydrocarbon haze on the early Earth was revived by Wolf and Toon (2010). Tian et al. (2005, 2006) argued that H escape from the early Earth was less efficient than previously believed and that a reducing atmosphere could be sustained longer than thought. Their conclusion depends on H escape being energy limited by a cold thermosphere on the early Earth instead of being diffusion limited as in Earth's present atmosphere.

Another approach to the problem of maintaining a reducing atmosphere is to find a sufficiently strong source of $NH_3$ (or $CH_4$) to balance the photochemical sink. Based on their experimental work, Brandes et al (1998) argue that mineral catalyzed reduction of $N_2$, nitrite, and nitrate to $NH_3$ can provide $5 \times 10^{17}$ mol $NH_3$ over $10^7 – 10^9$ years. The corresponding source strengths are $0.0085 – 0.85$ Tg $NH_3$ per year. However this is a small source of reduced nitrogen. Fegley et al. (1986) modeled HCN production from impact driven atmospheric shock heating and calculated a HCN deposition rate of ~ 8 – 38 Tg per year. Tian et al (2011) modeled photochemical production of HCN and calculated a HCN deposition rate of ~ 30 Tg per year. Also, modern day $NH_3$ emissions are ~ 54 Tg $yr^{-1}$, essentially all from anthropogenic and biogenic sources.

This chapter is organized as follows. We review the arguments for a secondary origin of the terrestrial atmosphere, i.e., by outgassing during and/or after accretion rather than by capture of solar nebula gas. Then we discuss sources of volatiles accreted by the Earth using meteorites as analogs for the material present in the solar nebula. The next section reviews heating during accretion of the Earth. Subsequently we describe chemistry of the silicate vapor, steam, and gaseous stages of atmospheric evolution on the early Earth. We close with a summary of the key questions that remain unresolved.

**Secondary origin of Earth's atmosphere**

Earth's atmosphere is secondary and originated by chemical reactions that released gases from volatile-bearing solids during and/or after its accretion. Earth's atmosphere is not primary, i.e., formed by capture of solar nebula gas during its formation. This conclusion is based upon the large depletions of the chemically inert noble gases (Ne, Ar, Kr, Xe) relative to chemically reactive volatiles such as H (as water), C, and N at Earth's surface. Helium is not considered in this comparison because it is continually escaping to space with an atmospheric lifetime of 0.9-1.8 Ma for [4]He and 0.4-0.8 Ma for [3]He (Torgersen 1989). Radon is not considered because its longest lived isotope [222]Rn decays with a half life of about 3.8 days. Also, when we talk about hydrogen we are concerned only with water and not with retention of hydrogen as $H_2$ gas.

Aston (1924a, b) discovered the enormous disparity between the abundances of the noble gases and other volatile elements. He plotted terrestrial elemental abundances as a function of mass up to a mass number of 142 and noted the "abnormal rarity of the inert gases" relative to that of the other elements. He concluded that "the earth has only one millionth part of its proper quota of the inert gases."

Using the improved astronomical and geochemical data accumulated during the intervening 25 years, Brown (1949) and Suess (1949) compared the cosmic abundances and Earth's near-surface inventories of H (as $H_2O$), C, N, Ne, Ar, Kr, and Xe. They concluded that all volatile elements are depleted on Earth relative to their solar abundances. However, the noble gases are much more depleted on Earth than chemically reactive volatiles such as H (as water), C, and N. The depletion is a chemical effect and is not due to physical effects such as escape.

Figure 1 shows a modern version of the argument made by Brown (1949) and Suess (1949). The depletion factors for the terrestrial abundances of the noble gases and the chemically reactive elements H (as water), C, N, F, S, and Cl are plotted on a logarithmic scale. The depletion factors are the terrestrial elemental abundance relative to silicon divided by the solar elemental abundance relative to silicon. For example, the Ne depletion factor is

$$D_{Ne} = \frac{(Ne/Si)_{Earth}}{(Ne/Si)_{solar}} \qquad (1)$$

The numerator is the Ne/Si mass ratio in the bulk silicate Earth and the denominator is the Ne/Si mass ratio in solar composition material. Similar equations give the depletion factors for all the other volatiles considered. Table 1 summarizes the solar and terrestrial elemental abundances used in the calculations.

The BSE in Table 1 is the bulk silicate Earth. It includes the atmosphere, biosphere, hydrosphere, and lithosphere. With the exception of sulfur, the BSE abundances are from Palme and O'Neill (2003). The selected sulfur abundance (124 µg/g) (Lodders and Fegley 2011) is smaller than the value of 200 µg/g from Palme and O'Neill (2003). However, our selected sulfur abundance agrees (within the uncertainty) with the value of $146 \pm 35$ µg/g for the MORB source region from Saal et al. (2002). For reference, published estimates of the sulfur abundance in the BSE range from $13 - 1000$ µg/g (Table 6.9 in Lodders and Fegley 1998). The solar abundances of Lodders et al. (2009) are used in Table 1 because they updated the abundances of C, N, O, F, Cl, S, and the noble gases. The "solar abundance" of water is calculated as the total abundance of oxygen minus the amount in rock, which is computed as the sum of oxygen in $SiO_2$ + MgO. The H/C mass ratio of 1.20 in the BSE (Palme and O'Neill 2003) is close (and identical within error) to the value of $0.99 \pm 0.42$ recommended by Hirschmann and Dasgupta (2009). Palme and O'Neill's selected H abundance in the BSE corresponds to about 3 times the mass of Earth's oceans, meaning about 2 oceans of water in the mantle. This is higher than other estimates which are closer to 2 times the mass of Earth's oceans, or about 1 ocean of water in the mantle (e.g., see Hirschmann and Dasgupta 2009; Saal et al. 2002). The atmospheric inventories of noble gases were taken as their BSE inventories. The values for argon refer to primordial $^{36+38}$Ar and exclude $^{40}$Ar, which is produced by radioactive decay of $^{40}$K in rocks and is the bulk (99.60%) of terrestrial argon. No corrections were made for the different isotopic compositions of solar and terrestrial Ne (e.g., $^{21}$Ne production from neutron capture and alpha particle emission by $^{24}$Mg in rocks), Kr, and Xe (e.g., $^{129}$Xe from decay of $^{129}$I, heavy isotopes from decay of $^{244}$Pu) because it does not change significantly the terrestrial inventories of these noble gases as is the case for Ar. (The same conclusions about large depletions of the noble gases relative to chemically reactive volatiles are drawn using the solar and terrestrial abundances of $^{20}$Ne, $^{84}$Kr, and $^{132}$Xe.)

As concluded by Brown (1949) and Suess (1949), Figure 1 shows that all volatiles – chemically reactive and inert – are depleted on Earth relative to their abundances in the solar nebula. The depletions are smallest for fluorine, chlorine, and sulfur; larger for H (as water), C, and N; and largest for Ne, Ar, Kr, and Xe. The reason for this is simple – chemically reactive volatiles were incorporated into minerals in the solid grains accreted by the Earth during its formation but the noble gases were not.

The differences in the depletion factors of the chemically reactive volatiles arise from their solar elemental abundances and the stability of different minerals in the grains accreted by the Earth.  For example, the solar elemental abundances of F, Cl, and S are small enough that these three elements can be completely incorporated into various minerals such as halite (NaCl), apatite [$Ca_5(PO_4)_3(F,Cl,OH)$], and  troilite (FeS) found in chondritic meteorites (see Table 2, which lists some of the volatile-bearing phases in chondrites). This is relevant because chondritic meteorites are relatively unaltered samples of solid material from the solar nebula and their composition is a guide to that of the solid grains accreted by the Earth.

However, mass balance prevents complete incorporation of H (as $H_2O$), C, and N into minerals formed with Mg, Si, Fe, and other rock-forming metals because the solar elemental abundances of H, C, and N are much larger than those of the rock forming metals. For example, the C/Si atomic abundance ratio is ~ 7.0 in solar composition material but only 0.76 in CI chondrites, the most volatile-rich meteorites. Likewise, the N/Si atomic abundance ratio is ~ 2.1 in solar composition material but only 0.055 in CI chondrites. Finally, the $H_2O$/Si molar abundance ratio is ~ 12.7 in solar composition material but only 5.1 in CI chondrites.

A small fraction of H (as $H_2O$), C, and N occurs in meteorites as organic matter, e.g., like the insoluble organic matter (IOM) in CI chondrites with a bulk composition of $C_{100}H_{72}N_2O_{14}S_4$ (Hayes 1967), carbides, carbonates, diamonds, nitrides, hydrous minerals, and dissolved in iron alloy. The less abundant, chemically reactive volatiles (S, Cl, and F) are less depleted than the more abundant, chemically reactive volatiles (H, C, and N) because the latter are incompletely incorporated into solids.

The chemically unreactive noble gases are even more depleted because they do not form minerals that occur in chondritic material, and by analogy in the solid grains accreted by the Earth. Only infinitesimal amounts of noble gases occur in chondrites where they are physically trapped in phases present in the meteorites.

The calculated depletion factors are uncertain for several reasons, but the uncertainties are much smaller than the large ($200 - 10^9$ times) differences between depletion factors of the noble gases and chemically reactive volatiles. Uncertainties include (1) taking the atmospheric inventory as the total inventory of noble gases on Earth, (2) using observed elemental abundances for the atmosphere, biosphere, hydrosphere, and part of the lithosphere (crust plus upper mantle) as the total terrestrial abundances of volatiles, (3) taking the chemical composition of the upper mantle as representative of the entire mantle, (4) the range of published values for the terrestrial abundance of Si, the normalizing element, and of the  chemically reactive volatile elements, and (5) uncertainties in solar elemental abundances. We address these uncertainties next.

(1) It is unlikely that significant amounts of Ne, Ar, Kr, or Xe are trapped inside the Earth. Arguing by analogy with nitrogen (with BSE abundance about twice the atmospheric inventory), the atmospheric inventories of the noble gases may be lower than the total terrestrial inventories by a factor of two. Geochemical modeling by Ballentine and Holland (2008) and by Palme and O'Neill (2003) shows only 45-50% of the BSE $^{40}$Ar inventory is in the atmosphere. Assuming equal degassing efficiencies for all noble gases means that the atmospheric inventories of $^{36+38}$Ar, Kr, and Xe may be only 50% of the total BSE inventories. Sanloup et al. (2005, 2011) suggested that at high pressure $XeO_2$ can substitute for $SiO_2$ in quartz and to lesser extent into olivine. Their proposals remain speculative although Brock and Schrobilgen (2011) synthesized $XeO_2$.

(2) Strictly speaking, the terrestrial abundances in Table 1 are lower limits and thus the depletion factors are upper limits, because we do not know elemental abundances in the Earth's core, although plausible estimates exist (McDonough 2003). If his estimates are correct, significant amounts of H (600 μg/g), C (2000 μg/g), N (75μg/g), S (19,000 μg/g), and Cl (200 μg/g) are in the outer core, and these elements are less depleted (more abundant) on Earth than shown in Table 1. It is unlikely that any noble gases are in the core. Xenon is the most likely candidate but Nishio-Hamane et al (2010) find no evidence for Xe alloy formation with or dissolution into Fe at core pressures.

(3) Palme and O'Neill (2003) consider whether the upper mantle is representative of the entire mantle and conclude that it is based on seismic tomography showing slab penetration into the lower mantle and the secular evolution of $\varepsilon_{Nd}$ and $\varepsilon_{Hf}$ to their present values in the depleted mantle reservoir.

(4) Published estimates for the BSE abundance of Si range from 20.80-23.3% (Table 6.9 of Lodders and Fegley 1998), with most estimates in the range of 21.0-21.6%. The Si isotopic composition of the BSE (Georg et al. 2007) is consistent with 21.22% Si in the BSE (Palme and O'Neill 2003), which we used in our calculations. However, uncertainties in the BSE abundance of Si affect all the calculated depletion factors equally. The range of values for abundances of chemically reactive volatiles in the BSE can be a factor of two or more (see Table 1). Sulfur is the worst case with estimated BSE abundances ranging from 13-1000 µg/g.

(5) As stated earlier, the solar elemental abundances of Lodders et al. (2009) are used because they updated the abundances of C, N, O, F, Cl, S, and the noble gases. If another solar abundance compilation were used (see the six other sets in Table 2.2 of Lodders and Fegley 1998), the order of magnitude differences between depletions of the noble gases and chemically reactive volatiles are still found.

As originally argued by Brown (1949), the depletions shown in Figure 1 demonstrate that Earth's atmosphere did not form by capture of solar nebula gas. If this had happened the $Ne/N_2$ abundance ratio of $2.3 \times 10^{-5}$ in Earth's atmosphere would be 134,000 times larger than actually observed. Neon and nitrogen have similar atomic weights and solar abundances so a physical process such as diffusion or hydrodynamic escape would not alter their abundance ratio significantly. Molecular nitrogen and $NH_3$, which are the major N-bearing gases in a solar composition atmosphere, have similar molecular weights to Ne. They should be affected by diffusive loss or hydrodynamic escape from a solar composition atmosphere to a similar extent as neon. Yet, Earth is much richer in nitrogen than neon. Likewise, argon and sulfur have similar solar abundances and capture of solar nebula gas would give a terrestrial $^{36+38}$Ar/S abundance ratio of ~ 0.25 (by mass) instead of $5 \times 10^{-8}$ as observed in the bulk silicate Earth. (This comparison and that in Figure 1 are based on $^{36+38}$Ar because $^{40}$Ar is produced by decay of $^{40}$K in rocks and was not present in the solar nebula.) Likewise, chlorine and argon have similar atomic weights and the molecular weight of HCl (the predicted Cl-bearing gas in the solar nebula) is even closer to that of Ar. However, the solar $^{36+38}$Ar /Cl mass ratio (0.055) is 275,000 times larger than the $^{36+38}$Ar /Cl mass ratio of $2 \times 10^{-7}$ in the BSE.

Starting with the discovery of $^3$He outgassing from mid-ocean ridges into seawater (Clarke et al. 1969; Lupton and Craig 1975), isotopic analyses of noble gases emitted from Earth's mantle suggested that Earth contains some noble gases that have "solar-like" isotopic compositions (Ne - Honda et al. 1991; Xe - Caffee et al. 1999). Originally these observations were taken as evidence of a primordial solar composition atmosphere on Earth (e.g., Porcelli et al. 2001 and references therein). However, more recent isotopic analyses of Ne, Ar, Kr, and Xe in $CO_2$ well gases by Ballentine and coworkers gives isotopic compositions consistent with solar wind implanted gas in meteorites (Ballentine and Holland 2008, Holland et al. 2009). Their results suggest the noble gases originated via outgassing of accreted material similar to average carbonaceous chondrites instead of from a captured solar composition atmosphere. It is also difficult to understand how $^3$He, with an atmospheric lifetime of 0.4-0.8 Ma, a low solubility in molten silicates, and a large vapor/melt partition coefficient, could be preserved on Earth since its formation.

**Source(s) of volatiles accreted by the Earth**

It is widely believed that the Earth accreted from a mixture of materials in the inner region of the solar nebula. For example, the two component models of Ringwood (1979) and Wänke (1981) postulate accretion of the Earth from varying proportions of oxidized, volatile-rich material similar to CI carbonaceous chondrites and reduced, volatile-poor material similar to (but with less volatiles than) enstatite chondrites. The equilibrium condensation – accretion model of Lewis (Barshay 1981; Lewis 1974, 1988) has the Earth and other terrestrial planets accreting material from feeding zones extending from inside to outside their present planetary orbits. More reduced, volatile-poor material came from the inner edge of the feeding zone and more oxidized, volatile-rich material came from the outer edge of the feeding zone. However, instead of using two explicitly defined components, chemical equilibrium calculations are used to define the composition of solid material across the planetary feeding zones (Barshay 1981). Anders and colleagues (e.g., Morgan and Anders 1979) theorized that the Earth (and other objects) formed from a mixture of chondritic materials that were subjected to the same cosmochemical processes as the chondritic meteorites (i.e., condensation, metal – silicate fractionation, volatile element loss, and so on). In these models, Earth's volatiles were provided by accretion of a late veneer of CV3 chondritic material. The oxygen isotope mixing (OIM) model uses the isotopic composition of oxygen to constrain the contributions of different types of chondritic material to the Earth and other rocky bodies (Lodders 1991, Lodders and Fegley 1997, Lodders 2000). Oxygen is used because it is the most abundant element in rock and is either the first or second (after iron) most abundant element in the terrestrial planets. The OIM model predicts Earth formed from a mixture of enstatite, ordinary, and carbonaceous chondritic material (70% EH, 21% H, 5% CV, and 4% CI). A large proportion (60-70%) of reduced, oxygen-poor material is also predicted by the heterogeneous accretion models of Rubie et al. (2011).

Working from a different starting point geochemists used chemical analyses of relatively unaltered rocks (fertile peridotites, spinel-lherzolites, and garnet lherzolites), petrological models of peridotite-basalt melting, and element ratios to determine the composition of the present day mantle and of the bulk silicate Earth (e.g., Jagoutz et al. 1979; Hart and Zindler 1986; Kargel and Lewis 1993; Allegre et al. 1995; Palme and O'Neill 2003). As a result of this work geochemists also concluded that the Earth has a kinship to chondritic meteorites, although it cannot be identified with any one chondrite group. Quoting Hart and Zindler (1986): "In other words, the Earth is not like any chondrite (in major elements) but is composed of its own blend of accretion products, with a bias toward a higher proportion of the high temperature refractory components (or equivalently, a lower proportion of the volatiles and partially refractory components)."

At the time these models were done it was thought that igneous differentiation of meteoritic material took place after accretion of the Earth was complete. However, dating by three relative chronometers ($^{182}$Hf – $^{182}$W, $^{26}$Al – $^{26}$Mg, and $^{53}$Mn – $^{53}$Cr) indicates that the HED meteorite parent body differentiated during the first 1-10 Ma of solar system history (e.g., Jacobsen et al. 2008, Trinquier et al 2008, Schiller et al 2011). The HED (howardite, eucrite, diogenite) meteorites are a major group of igneously differentiated (achondritic) meteorites. Lead – lead ($^{207}Pb - ^{206}Pb$), $^{26}Al - ^{26}Mg$, and $^{53}Mn - ^{53}Cr$ dating also indicate that the parent body of the Asuka 881394 basaltic meteorite formed and differentiated within 3 Ma of solar system formation (Wadhwa et al 2009). Consequently, the Earth plausibly accreted a mixture of chondritic and achondritic material during its formation. Thus it is possible that achondritic material like the HED meteorites also was a volatile source for the Earth.

Figure 2 compares the $H_2O$, C, and N contents of the three major classes of chondrites (ordinary, enstatite, and carbonaceous) with those of the HED meteorites and the bulk silicate Earth. The data in Figure 2 are for the Orgueil CI chondrite (Lodders et al. 2009), average ordinary chondrites (Schaefer and Fegley 2007), average enstatite chondrites (C and N from Lodders and Fegley 1998, water in the Hvittis enstatite chondrite from Robert et al. 1987), and HED meteorites (Grady and Wright 2003). There are no analyses for water in HED meteorites. These references and Lodders (2003) describe the analytical data in more detail and we refer the reader to these papers.

Figure 2 shows that chondritic material contains more $H_2O$, C, and N than the Earth, and the HED achondrites contain less C and N than the Earth. However, other differentiated meteorites contain more volatiles. The SNC meteorites, believed to come from Mars, contain several hundred μg/g water and carbon (Lodders 1998). But the high volatile content of the SNC meteorites is inherited from their planetary parent body. The ureilites, which are carbon-rich achondrites, contain up to 7 wt % carbon (primarily as graphite and/or diamond) and up to 30 μg/g nitrogen (associated with graphite or diamond) (Grady and Wright 2003). But ureilites are relatively rare (15% of all achondrites) and are anhydrous. The available data on water, C, and N in achondrites shows it is unlikely igneously differentiated material provided significant amounts of volatiles to the Earth. In any case, volatiles in Earth-forming materials were outgassed by heating during accretion, which is discussed next.

**Heating during accretion of the Earth**

Heating during accretion of the Earth can be substantial. Overall, the thermal energy (E) released by the accretion of the Earth is

$$E \sim \frac{GM_P^2}{R_P} \sim 4 \times 10^{32} \, J \tag{2}$$

where G is the universal constant of gravitation ($6.6726 \times 10^{-11}$ $m^3 \, kg^{-1} \, s^{-2}$), $M_P$ is the planetary mass in kilograms, $R_P$ is the planetary radius in kilometers, and the thermal energy E is in Joules. For example, accretion of the Earth ($M_P \sim 6 \times 10^{24}$ kg, radius $R_P \sim$ 6370 km) releases $4 \times 10^{32}$ J of heat. This is more than enough energy to heat up and vaporize the bulk silicate Earth if it were all released at one time.

As discussed earlier, the BSE includes the atmosphere, biosphere, hydrosphere, and lithosphere. It has a mass of $M_{BSE} \sim 4.03 \times 10^{24}$ kg, dominated by the mantle (4.007 $\times 10^{24}$ kg, 99.4% of the BSE). The mean specific heat of rock having the composition of the rocky part of the bulk silicate Earth is 1.16 J/g K, or about $4.7 \times 10^{27}$ J/K for the entire BSE. For comparison, Solomatov (2007) gives 1 J/g K for the specific heat of a magma ocean (composed of olivine, orthopyroxene, and clinopyroxene) on the early Earth and Sleep et al. (2001) use a value of 1.05 J/g K for the specific heat of the core plus mantle. Thus taking only the specific heat of rock into account, the heat released by accretion of the Earth could heat the BSE to ~ 85,000 K.

A more exact calculation uses the thermodynamic properties of forsterite and also considers the energy needed to melt rock, vaporize the melt, and heat the rock vapor. To first approximation, the mantle (and thus the BSE) is mainly composed of forsterite $Mg_2SiO_4$ ($\mu \sim 140.7$ g mol$^{-1}$). The energy required to heat up, melt, and vaporize one mole of forsterite from room temperature (298 K) to its one bar boiling point ($\sim 3690$ K) is about 1,186 kilojoules. This is composed of four terms: (1) the enthalpy for heating forsterite from 298 K to its melting point of 2163 K ($H_{2163} - H_{298} \sim 330.0$ kJ), (2) the enthalpy for melting solid forsterite ($\Delta_{fus}H \sim 114 \pm 20$ kJ), (3) the enthalpy for heating molten forsterite from the melting point to the one bar boiling point ($H_{3690} - H_{2163} \sim 313$ kJ), and (4) the enthalpy of vaporization of molten $Mg_2SiO_4$ ($\sim 429 \pm 40$ kJ). The enthalpy of vaporization includes the energy needed to dissociate molten $Mg_2SiO_4$ into the mixture of gases (Mg, $O_2$, SiO, O, MgO, $SiO_2$) that constitute "forsterite" vapor.

The enthalpy terms for heating solid and molten forsterite are from the JANAF Tables (Chase 1998), the enthalpy of melting is the value of Navrotsky et al. (1989) measured at 1773 K and recalculated to the melting point (2163 K) by Richet et al. (1993), and the enthalpy of sublimation ($543 \pm 34$ kJ) is the value of Nagahara et al. (1994). The enthalpy of vaporization ($\Delta_{vap}H$) is $429 \pm 40$ kJ and is the sublimation enthalpy minus the melting enthalpy. The one bar boiling point of $\sim 3690$ K is calculated from the triple point pressure ($5.2 \times 10^{-5}$ bar) and enthalpy of vaporization by integrating the Clausius – Clapeyron equation between the triple point ($T_1$) and the boiling point ($T_2$) assuming a constant $\Delta_{vap}H$ and rearranging to solve for $T_2$ (pp. 250-256 in Fegley 2013)

$$\int d\ln P = \int \frac{\Delta H}{RT^2}\,dT = \frac{\Delta H}{R}\int \frac{1}{T^2}\,dT \qquad (3)$$

$$\frac{1}{T_2} = -\frac{R}{\Delta_{vap}H}\ln(\frac{P_2}{P_1}) + \frac{1}{T_1} \tag{4}$$

A more sophisticated approach (using a constant $\Delta C_P$ for the vaporization reaction and thus a three term vapor pressure equation) is unwarranted given the large uncertainties in the phase transition enthalpies. The derived vapor pressure equation (2163 – 3690 K) for molten forsterite is

$$\log P_{vap}\left(Mg_2SiO_4, liq\right) = 6.08 - \frac{22,409}{T} \tag{5}$$

Equation (5) gives the total pressure (bar) of the saturated vapor in equilibrium with molten forsterite. As noted above, forsterite vapor is a mixture of several gases (Mg, $O_2$, SiO, O, MgO, $SiO_2$). The relative abundance of the different gases in forsterite vapor depends upon temperature but the bulk composition of the vapor is the same as that of forsterite, i.e., $Mg_2SiO_4$ vaporizes congruently.

Based on our calculations using thermodynamic properties of forsterite, the enthalpy ($H_{vap}$) to vaporize the bulk silicate Earth is

$$H_{vap} = \frac{M_L}{\mu} \times 1186 = \frac{4.007 \times 10^{27}\,g}{140.7\,g\,mol^{-1}} \times 1.186 \times 10^6\,J\,mol^{-1} \sim 3.4 \times 10^{31}\,J \tag{6}$$

This is about 8.4% of the thermal energy released by accretion of the Earth (according to Eq. 2), so vaporization of the BSE could occur if all energy were released at one time.

This calculation used pure forsterite ($Fo_{100}$) instead of the forsterite-rich olivine ($Fo_{90}Fa_{10}$) found in Earth's mantle. However, pure fayalite has a smaller enthalpy of fusion ($89.3 \pm 1.1$ kJ $mol^{-1}$) and smaller enthalpy of sublimation ($502 \pm 9$ kJ $mol^{-1}$) than pure forsterite (Stebbins and Carmichael 1984, Nagahara et al. 1994). The calculations with pure forsterite over estimate the energy required for melting and vaporization.

Vaporization of the Earth's core requires a similar amount of energy to that needed to vaporize the BSE. Data in the JANAF Tables (Chase 1998) show the energy required to heat up, melt, and vaporize one mole of iron metal from room temperature (298 K) to its one bar boiling point (3133 K) is about 483 kilojoules. This is composed of four terms: (1) the enthalpy for heating iron metal from room temperature to its melting point ($H_{1809} - H_{298} \sim 58.6$ kJ), (2) the enthalpy for melting iron metal ($\sim 13.8$ kJ), (3) the enthalpy for heating molten iron from its melting point to its boiling point ($H_{3133} - H_{1809} \sim 60.95$ kJ), and (4) the enthalpy of vaporization of molten iron at its boiling point ($\sim 349.58$ kJ). Iron vapor is monatomic Fe gas. The mass of Earth's core (taken as pure iron for the calculations) is $1.94 \times 10^{24}$ kg, which is about $3.48 \times 10^{25}$ moles of iron. Heating up, melting, and vaporizing the Fe core requires about $1.7 \times 10^{31}$ J. Thus, it takes about $5.1 \times 10^{31}$ J to heat the entire Earth from 298 K and vaporize it. This is $\sim 13\%$ of the energy released by Earth's accretion.

Hafnium – tungsten dating shows accretion of the Earth was rapid with a mean accretion time of 10 million years and a total time of 30 million years (after the formation of the solar system 4.568 Ga ago) (e.g., Jacobsen et al. 2008). The average accretion rate during the 30 Ma period was about $2 \times 10^{17}$ kg per year (about twice the mass of the asteroid 253 Mathilde), and plausibly decreased from higher to lower values as Earth grew. Objects accreted by the Earth had impact velocities of about 10 km s$^{-1}$. Thus, the kinetic energy ($E_K$ in Joules) from each impact is

$$E_K = \frac{1}{2}Mv^2 = 5 \times 10^7 M \qquad (7)$$

The impactor mass (M) is in kilograms in this equation. The average kinetic energy input during the 30 Ma period for accretion of the Earth is about $10^{25}$ J per year. As shown in Table 3, the average kinetic energy input per year is less energy than delivered by the impact of a Pallas-size planetesimal. As a consequence the global thermal effects for the proto-Earth were plausibly limited to boiling the oceans (if any) and heating the resulting vapor to 2000 K. Table 3 shows that larger impacts are needed for larger thermal effects.

Computer models of terrestrial planet accretion predict that the Earth formed from a relatively small number of large planetesimals and that impacts between large objects were frequent in the early solar system (e.g., see Canup 2008 and references therein). For example, the impact of a Mars size object ($\sim 0.1 \ M_E$) with the proto-Earth ($\sim 0.9 \ M_E$) at or near the end of Earth's accretion is thought to have formed the Moon (Hartmann and Davis 1975, Cameron and Ward 1976). Table 3 shows the energy delivered by the impact of a Mars size object is $\sim 3 \times 10^{31}$ J, which is sufficient to vaporize the bulk silicate Earth. Radiative cooling time estimates (flux $= \sigma T^4$) suggest the silicate atmosphere on the post-impact Earth existed for $10^3 - 10^4$ years (Stevenson 1987, Thompson and Stevenson 1988, Pahlevan and Stevenson 2007, Sleep et al. 2001, Sleep 2010) depending on the effective radiating temperature.

**Earth's silicate vapor atmosphere**

The hydrodynamic simulations of the Moon-forming impact show that the post-impact Earth is heated to many thousands of degrees. For example, the surface is heated above 6000 K and the interior above 15,000 K (Figures 3 and 4 of Canup 2008). These very high temperatures are well above the melting points and boiling points of rocks and minerals. Thoria ($ThO_2$), the major constituent of natural thorianite and the highest melting point oxide, melts at $3640 \pm 30$ K and boils at $\sim 4300$ K (Ackermann et al. 1963; Rand 1975). The melting points of more common rocks (e.g., 1973 K for dry peridotite from Takahashi 1986) and minerals (forsterite 2163 K, enstatite 1830 K, anorthite 1830 K, and spinel 2408 K) are lower than that of thoria. Furthermore water depresses the melting points of dry peridotite and other rocks and minerals. Thus, the temperature of Earth's surface after the Moon-forming impact is thousands of degrees higher than the liquidus temperature of peridotite or any other rock or mineral.

The experimentally measured boiling point (one atmosphere) of tektite glass, which is about 70% silica, is ~ 3100 K (Centolanzi and Chapman 1966). This agrees fairly well with the calculated one bar boiling point of 3157 K for molten silica. We computed this by considering chemical equilibria between molten silica and all species in the saturated vapor (SiO, $O_2$, O, $SiO_2$, Si) using the MAGMA code (Fegley and Cameron 1987, Schaefer and Fegley 2004). Other computed values for the boiling point of molten silica are 3070 (Schick 1960), 3143 K (Krieger 1965), and 3157 K (Melosh 2007). The vapor pressure equation for molten silica (2000 – 6000 K) is

$$\log P_{vap}(SiO_2, liq) = 8.203 - \frac{25{,}898.9}{T} \tag{8}$$

Equation (8) gives the total pressure (bar) of the saturated vapor in equilibrium with molten silica. Silica vaporizes congruently, i.e., the saturated vapor has the bulk composition $SiO_2$, but the abundances of the different species in the vapor (SiO, $O_2$, O, $SiO_2$, Si) vary with temperature. Previously, several groups modeled the silicate vapor atmosphere produced by the Moon-forming impact as pure silica vapor (e.g., Stevenson 1987, Thompson and Stevenson 1988, Ward 2012) but this is only a rough approximation because neither the Earth nor the Moon are pure silica.

As discussed earlier, the calculated one bar boiling point of molten forsterite is ~ 3690 K. We used the enthalpy of fusion (89.3 ± 1.1 kJ mol$^{-1}$) from Stebbins and Carmichael (1984) and the enthalpy of sublimation ($502 \pm 9$ kJ mol$^{-1}$) of fayalite from Nagahara et al. (1994) to compute the enthalpy of vaporization ($412.7 \pm 9.1$ kJ mol$^{-1}$) of molten fayalite at its triple point. Combining this value with the triple point pressure of $6.3 \times 10^{-8}$ bar from Nagahara et al. (1994) and integrating the Clausius – Clapeyron equation with a constant enthalpy of vaporization give a one bar boiling point of ~ 2965 K and the vapor pressure equation (1490 – 2965 K)

$$\log P_{vap}(Fe_2SiO_4, liq) = 7.27 - \frac{21,558}{T} \tag{9}$$

Again, the more sophisticated approach (using a constant $\Delta C_P$ for the vaporization reaction and thus a three term vapor pressure equation) is unwarranted given the large uncertainty in the vaporization enthalpy. Equation (9) gives the total pressure (bar) of the saturated vapor in equilibrium with molten fayalite. Fayalite vaporizes congruently like forsterite. Thus fayalite vapor has the same bulk composition as fayalite ($Fe_2SiO_4$), but the vapor is a mixture of gases, e.g., Fe, $O_2$, O, SiO, FeO, $SiO_2$ whose relative abundance varies with temperature.

Thermodynamic data on the vaporization of molten Fe from the JANAF Tables give a one bar boiling point of 3133 K and the vapor pressure equation (1809 – 4000 K)

$$\log P_{vap}(Fe, liq) = 15.4345 - \frac{21,596.52}{T} - 2.4432 \log T \tag{10}$$

The three term equation is warranted for the vapor pressure (bar) of molten Fe because the data are more extensive and accurate.

Based on the hydrocode simulations and the discussion above, we explicitly assume that the surface of the post-impact Earth is molten, at least until the surface temperature decreases to ~ 2000 K. Atmospheric pressure at the Earth's surface is thus given by the saturated vapor pressure of the molten rock plus the partial pressures of volatile gases (e.g., $H_2O$, $CO_2$) that partition between the molten silicate and its vapor. The magma ocean and the overlying atmosphere are expected to rapidly equilibrate on a timescale much shorter than either the lifetime of the silicate vapor atmosphere ($10^3 - 10^4$ years) or the crystallization time ($10^4 - 10^7$ years) of the magma ocean (e.g., see Hirschmann 2012). The presence (if any) of a solid "skin" on top of the molten silicate should not change our conclusion that the atmospheric pressure is due to the sum of these two terms. The two reasons for this are (1) that it is unlikely that a solid "skin" would hermetically seal the magma from the vapor and (2) the Gibbs energy difference, and hence the vapor pressure difference, between the solid and its melt are small in the vicinity of the melting point. In the case of forsterite (for which data are available in the JANAF Tables), the melting reaction and its standard Gibbs energy change ($\Delta G^o$, in J mol$^{-1}$) are

$$Mg_2SiO_4 \text{ (forsterite)} = Mg_2SiO_4 \text{ (liquid)} \qquad (11)$$

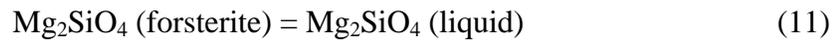

$$\Delta G^o = 70,903.14 - 32.78T \qquad (12)$$

Equation (12) shows the difference in the Gibbs energies of solid and molten forsterite is < 33 J mol$^{-1}$ K$^{-1}$ in the vicinity of its melting point (2163 K). The corresponding change in vapor pressure is ~ 2% for a solid skin which is 10 K cooler than the underlying magma and ~ 20% for a solid skin which is 100 K cooler. The change in the solubility of $H_2O$, $CO_2$, and other gases in the molten silicate is probably also small for such small temperature changes.

The molten rock has the composition of the bulk silicate Earth (see Table 4). Takahashi (1986) measured melting of the dry, fertile peridotite KLB-1 and found a liquidus temperature of ~ 1700 C (1973 K) at one bar. We take his value as the liquidus temperature for the molten bulk silicate Earth. For comparison Sleep et al. (2001) used a liquidus temperature of 2036 K. We calculated the saturated vapor pressure and vapor composition over the molten BSE magma from 2000 – 6000 K using the MAGMA code. The vapor pressure equation is

$$\log P_{vap}(BSE\,magma) = -30.6757 - \frac{8,228.146}{T} + 9.3974 \log T \tag{13}$$

Equation (13) gives the total pressure (bar) of the saturated silicate vapor in equilibrium with the molten BSE magma. The calculated one bar boiling point of the molten BSE magma is ~3352 K, lower than that of pure forsterite because of the presence of the more volatile elements Na, K, and Fe in the BSE magma. We discuss the vapor composition below. We showed the good agreement of the MAGMA code with experimental data and other computer codes in two prior papers (Fegley and Cameron 1987, Schaefer and Fegley 2004) and we refer the reader to them.

Figure 3 shows the vapor pressure of the molten BSE magma and compares it to that of molten forsterite, silica, fayalite, and iron metal. The vapor pressure of molten forsterite is lower than that of the other materials, which is consistent with it having the highest melting point and lowest content of volatile elements. The BSE magma has a vapor pressure intermediate between that of fayalite (Fa) and forsterite (Fo), which is qualitatively consistent with the olivine composition of Earth's mantle which is about $Fo_{90}Fa_{10}$. However, the vapor pressure of the BSE magma is higher than that of Fo-rich olivine because of the presence of Na and K, which are very volatile at magmatic temperatures and are easily vaporized. This is shown for example, by mass spectrometric studies of heated lunar and terrestrial basalts and the Holbrook ordinary (L6) chondrite (DeMaria et al. 1971, DeMaria and Piacente 1973, Naughton et al. 1971, Gooding and Muenow 1976, 1977, Markova et al. 1986).

Figure 4 shows the composition of the saturated vapor in equilibrium with the BSE magma. The four major gases in the vapor over the 2000 – 6000 K temperature range are Na, SiO, $O_2$, and O. These four gases comprise over 90% of the saturated silicate vapor. Diatomic and monatomic oxygen comprise 21 – 36% of the vapor depending on the temperature. The silicate vapor is orders of magnitude more oxidizing (i.e., has a larger oxygen fugacity $fO_2$) than $H_2$-rich solar composition gas.

All other species are less abundant than Na, SiO, $O_2$ and O in the silicate vapor. For example, at 6000 K, the next two most abundant gases are MgO (~ 8%) and $SiO_2$ (~ 6%). Potassium is also volatile but monatomic K and other K-bearing gases are trace species in the vapor because of (1) the low K abundance in the BSE (Na/K atomic ratio ~ 17 from Table 4) and (2) preferential partitioning of Na relative to K into the vapor, e.g., Na/K atomic ratio ~ 6,300 at 2000 K in Figure 4 (also see the discussion in Schaefer and Fegley 2004). Although Mg is about as abundant as Si in the BSE, the two major Mg-bearing gases (Mg and MgO) are much less abundant than SiO (the major Si-bearing gas) because MgO has a much lower vapor pressure than does silica, e.g., at 2020 K the saturated vapor pressures of silica and MgO are $2.4 \times 10^{-5}$ bar and $1.2 \times 10^{-7}$ bar (Farber and Srivastava 1976a), respectively. Conversely FeO is more volatile than MgO, and the Fe/Mg and FeO/MgO ratios in the vapor are greater than unity below ~ 2600 K. The opposite is true at higher temperatures. Alumina, CaO, and $TiO_2$ have much lower vapor pressures than the other oxides and their gases are generally much less abundant than those of the other oxides.

Knudsen effusion mass spectroscopy (KEMS) studies of solid and molten oxide vaporization are important for understanding the composition of Earth's silicate vapor atmosphere. The abundances of O and $O_2$ are coupled to each other via the equilibrium

$$2 \text{ O (g)} = O_2 \text{ (g)} \tag{14}$$

Lower temperatures and higher pressures favor $O_2$ while higher temperatures and lower pressures favor monatomic O. Silicon monoxide SiO, $O_2$, and O are the major species in the saturated vapor over solid and molten silica (e.g., see Firsova and Nesmeyanov 1960, Kazenas et al. 1985, Nagai et al. 1973, Shornikov et al. 1998, Zmbov et al. 1973). The abundances of SiO, $SiO_2$, which is less abundant, and Si, which is much less abundant, are coupled to one another via the equilibria

$$\text{SiO (g)} + \text{O (g)} = SiO_2 \text{ (g)} \tag{15}$$

$$\text{SiO (g)} = \text{Si (g)} + \text{O (g)} \tag{16}$$

Monatomic Mg and $O_2$ are the major species over solid MgO and MgO gas is much less abundant (e.g., see Farber and Srivastava 1976a; Kazenas et al. 1983; Porter et al. 1955). Monatomic Ca and $O_2$ are the major species over solid CaO and CaO (g) is much less abundant (Farber and Srivastava 1976b, Samoilova and Kazenas 1995). Alumina vaporization is more complex and occurs via a combination of dissociation to the elements and the production of various Al oxides (e.g., Chervonnyi et al. 1977, Drowart et al. 1960, Farber et al. 1972, Ho and Burns 1980). Saturated alumina vapor contains Al, O, $O_2$, AlO, $Al_2O$, $Al_2O_2$, and $AlO_2$. The O (g) partial pressure in the BSE saturated vapor is sufficiently high that AlO is the major Al-bearing gas due to the equilibrium

$$\text{Al (g)} + \text{O (g)} = \text{AlO (g)} \tag{17}$$

Gas phase equilibria in the silicate vapor atmosphere are rapid because of the high temperatures and pressures. This leads to chemical and isotopic equilibrium, e.g., oxygen isotopic equilibrium within very short times. The elementary reactions below convert monatomic and monoxide gases of several of the major metals in the silicate vapor

$$\text{Si} + \text{O}_2 \rightarrow \text{SiO} + \text{O} \tag{18}$$

$$k = 1.72 \times 10^{-10}(T/300)^{-0.53}\exp(-17/T) \;\; \text{cm}^3 \text{ s}^{-1} \tag{19}$$

$$\text{Fe} + \text{O}_2 \rightarrow \text{FeO} + \text{O} \tag{20}$$

$$k = 2.09 \times 10^{-10}\exp(-10{,}175/T) \;\; \text{cm}^3 \text{ s}^{-1} \tag{21}$$

$$\text{Mg} + \text{O}_2 \rightarrow \text{MgO} + \text{O} \tag{22}$$

$$k = 4.81 \times 10^{-10}\exp(-8551/T) \;\; \text{cm}^3 \text{ s}^{-1} \tag{23}$$

Similar reactions can be written for the other metals in the silicate vapor atmosphere. The bimolecular reaction rate constants are from LePicard et al. (2001) for Si and the NIST chemical kinetic database for Fe and Mg. The chemical lifetimes ($t_{chem}$, in seconds) for Si, Fe, and Mg are given by the equation

$$t_{chem}(i) = \frac{1}{[O_2]k_i} \tag{24}$$

The $k_i$ is the reaction rate constant for the appropriate metal atom (Si, Fe, or Mg). The $[O_2]$ is the molecular number density of $O_2$ at any given temperature from our chemical equilibrium calculations and is given by

$$[O_2] = \frac{P_{O_2}N_A}{RT} \tag{25}$$

In Eq. (25), $N_A$ is Avogadro's number, R is the ideal gas constant, T is temperature (K), and $P_{O_2}$ is the $O_2$ partial pressure in the saturated silicate vapor.

At 2000 K, the chemical lifetimes of Si, Mg, and Fe are $10^{-3.90}$, $10^{-2.93}$, and $10^{-2.21}$ seconds, respectively. Reaction rates increase exponentially with temperature (as shown by Eqs. 19, 21, and 23) and the chemical lifetimes of Si, Mg, and Fe are even smaller at higher temperatures. Thus chemical and oxygen isotopic equilibrium is established very rapidly in Earth's silicate vapor atmosphere (also see Pahlevan et al. 2011 in this regard).

Chemical and oxygen isotopic equilibrium also plausibly occurs in the proto-lunar disk of molten silicate and vapor and may explain the very similar oxygen isotopic compositions of the Earth and Moon (Wiechert et al. 2001). For example, at 2000 K, the $O_2$ molecular number density can drop by a factor of $10^6$ and the chemical lifetimes given by Eq. (24) are still ~ one hour (Fegley et al. 2012). However, thermal escape of volatile elements and compounds from the proto-lunar disk may also have occurred. High precision Zn isotopic measurements of lunar basalts by Paniello et al. (2012) show that lunar basaltic rocks are enriched in heavier Zn isotopes and have lower Zn concentrations relative to the Earth. Their results indicate evaporative Zn loss during lunar formation.

Figure 3 also shows a line giving the total pressure of $H_2O$, C- and S-bearing gases. This is computed assuming all of the $H_2O$, carbon, and sulfur in the bulk silicate Earth (Table 1), which are $4.32 \times 10^{21}$, $4.03 \times 10^{20}$, and $5.00 \times 10^{20}$ kg, respectively, are in the atmosphere. In this case, the pressures (P = m/g) of $H_2O$, C- and S-bearing gases are ~ 820, 80, and 96 bars, respectively, summing to ~ 1000 bars. (The BSE carbon abundance in Palme and O'Neill (2003) is from Zhang and Zindler (1993). If we follow Zhang and Zindler (1993) and assume that all carbon in the BSE is carbonate, the $CO_2$ pressure would be ~ 250 bars instead of ~ 80 bars.) All three volatiles are soluble to varying extents in silicate melts, with water and sulfur being more soluble than C-bearing gases ($CO_2$, $CO$, $CH_4$). Holding other variables constant, sulfur solubility depends upon oxygen fugacity and has a minimum at about the $fO_2$ of the Ni-NiO buffer. Sulfur dissolves as sulfide at lower $fO_2$ and as sulfate at higher $fO_2$ and has solubilities in the $100 – 10,000$ µg/g range depending on melt composition, P, T, and $fO_2$ (Baker and Moretti 2011).

Experimental studies of water and $CO_2$ solubility in different types of silicate melts (e.g., basaltic, feldspathic, granitic, haploandesitic, pegmatitic, rhyolitic, silicic) show that their solubility increases with increasing pressure and decreases slightly (retrograde solubility) with increasing temperature (e.g., Burnham and Jahns 1962, Goranson 1931, Hamilton et al. 1964, Holz et al. 1995, Karsten et al 1982, Mysen and Wheeler 2000, Schmidt and Behrens 2008, Yamashita 1999). However, the temperature coefficients are small, e.g., ~ $0.1 – 0.3$ wt. % $H_2O$ per 100 K increase and have been measured over fairly small temperature intervals of 300 K or less at temperatures at or below 1300 C (1573 K). A molecular dynamics simulation of $CO_2$ solubility in silicate melts covers the $1200 – 2000$ C ($1473 – 2273$ K) range, and predicts decreasing solubility with increasing temperature (Guillot and Sator 2011). But there are no experimental data for $H_2O$, $CO_2$, or sulfur solubility in molten peridotite or other silicate melts at the temperatures expected on Earth's surface after the Moon-forming impact.

We modeled $H_2O$ and $CO_2$ solubility in silicate melts using the VolatileCalc program of Newman and Lowenstern (2002). The calculated solubilities of $H_2O$ (820 bar) and $CO_2$ (80 bar) at 2000 K are 3.06 wt. % and 36 µg/g, respectively. VolatileCalc predicted prograde solubility for water with a temperature coefficient of 0.029 wt. % per 100 degrees and a constant solubility for $CO_2$. We did not model sulfur solubility because of the complex dependence on T, P, $fO_2$, and melt composition.

The line for these volatiles in Figure 3 may thus be an upper limit, but a comparison with the vapor pressure curves of the molten silicates is instructive. Figure 3 shows that the atmospheric partial pressure of water and other volatiles is larger than the saturated vapor pressure of the BSE magma until ~ 5525 K. If 90% of $H_2O$, C, and S are dissolved in the magma, 100 bar of volatiles remains in the atmosphere and the magma must be at ~ 4630 K for its vapor pressure to be comparable. Equality of the molten silicate vapor pressure and volatile partial pressure occurs at 3925 K if 99% of $H_2O$, C, and S are dissolved in the magma, and 10 bar of volatiles remain. Even if 99.9% of $H_2O$, C, and S are dissolved in the magma and only 0.1% remains in the atmosphere ($P_{volatile} = 1$ bar), the BSE magma must be at 3352 K to have the same pressure. In other words, the silicate vapor atmosphere may be mainly steam, C-, and S-bearing gases even if most of the water, C, and S in the BSE is dissolved in the super-liquidus BSE magma. At present the partitioning of volatile elements between the magma and silicate vapor cannot be computed with certainty and new experimental data on the solubility of water, C-, and S-bearing gases in silicate melts, especially molten peridotite, at super-liquidus temperatures are required. This is important to do because the most abundant gases (Na, O, $O_2$) in the saturated silicate vapor are infrared inactive. Silicon monoxide has strong IR bands at 4 μm (first overtone) and 8 μm (fundamental) and lines throughout the millimeter region. By analogy with $CO_2$, $SiO_2$ (with a mole fraction > 0.1% for T > 2300 K) may be a good greenhouse gas. However the most efficient greenhouse gases in the silicate vapor atmosphere may be $H_2O$, $CO_2$, and $SO_2$.

Fegley and Schaefer (2006) modeled condensate cloud formation in Earth's silicate vapor atmosphere assuming an adiabatic temperature gradient and a surface temperature of 2000 K. As Figure 3 shows, the silicate vapor is dominantly Na, $O_2$, O with smaller amounts of Fe, SiO, FeO, Mg, and other gases at this temperature. Their predicted cloud condensation sequence as a function of altitude in the silicate vapor atmosphere is $Mg_2SiO_4$ (1955 K, 4.5 km), liquid $CaAl_2Si_2O_8$ (1931 K, 7 km), liquid $CaSiO_3$ (1896 K, 9.5 km), cristobalite (1870 K, 13 km), $Fe_3O_4$ (1817 K, 18 km), rutile (1602 K, 40 km), and $Na_2O$ (1169 K, 82 km). The $Na_2O$ cloud was the most massive, followed by the cristobalite, magnetite, and forsterite cloud layers. The liquid anorthite, liquid wollastonite, and rutile layers were thin hazes because their abundances are severely limited by the small amounts of Ca, Al, and Ti in the silicate vapor at 2000 K. Monatomic K remained in the gas and never condensed from the atmosphere. A different condensate cloud sequence with liquid oxide and silicate clouds was predicted for a hotter atmosphere (5000 K surface T). This is consistent with the temperature-dependent composition of the silicate vapor (Fig. 3).

In principle, the results of Fegley and Schaefer (2006) can be checked by spectroscopic observations of hot rocky exoplanets. The three known examples as of the time of writing (Sept 2012) are CoRoT-7b (sub-stellar surface T ~ 2474 K, Léger et al. 2011), Kepler-10b (sub-stellar surface T ~ 3038 K, Léger et al 2011), and 55 Cnc e (sub-stellar T ~ 2800 K, Winn et al. 2011). All three planets are probably tidally locked with the same side always facing the parent star. The average surface temperatures are lower than the sub-stellar temperatures and are ~ 1810 K (CoRoT-7b), ~1833 K (Kepler-10b), and ~1967 K (55 Cnc e).

To date the only published spectroscopic observations of hot rocky exoplanets are those of Guenther et al. (2011) who searched for gaseous Ca, $Ca^+$, CaO, and Na in the exosphere of CoRoT-7b. They did not detect any of these gases. However, all Ca and Na are predicted to condense into clouds below 100 km altitude and to be absent in the exospheres of hot rocky exoplanets (Fegley and Schaefer 2006, Schaefer and Fegley 2009, Schaefer et al 2012).

Fegley and Schaefer (2006) did not model photochemistry of the silicate vapor atmosphere but some general statements can be made. Nemtchinov et al. (1997) give linear absorption coefficients for H chondrite vapor in the visible and IR ranges. At one bar pressure the optical absorption coefficients are $0.01 - 10$ $cm^{-1}$ and the IR absorption coefficients are $10^{-6} - 1$ $cm^{-1}$. Much of the opacity is due to Fe-bearing gases. Iron is more abundant in H chondrites than in the BSE (16.7 wt. % Fe metal vs. 8.18 % FeO) and the opacity of BSE vapor may be lower. Silicon monoxide absorbs light shortward of 300 nm (Podmoshenskii et al. 1968, Vorypaev 1981, Matveev 1986)

$$SiO + h\nu \rightarrow Si + O \quad \lambda < 300 \text{ nm} \qquad (26)$$

However reaction (18) may rapidly regenerate SiO gas. Solar UV photolysis of $O_2$ ($\lambda < 240$ nm) may produce monatomic oxygen and ozone via the Chapman cycle:

Photolysis of $O_2$ $\qquad O_2 + h\nu \rightarrow O + O$ ($\lambda = 180 - 240$ nm) $\qquad (27)$

Ozone production $\qquad O + O_2 + M \rightarrow O_3 + M$ $\qquad (28)$

Ozone photolysis $\qquad O_3 + h\nu \rightarrow O(^1D) + O_2$ ($\lambda = 200 - 300$ nm) $\quad (29)$

Oxygen production $\qquad O + O + M \rightarrow O_2 + M$ $\qquad (30)$

Ozone destruction $\qquad O + O_3 \rightarrow O_2 + O_2$ $\qquad (31)$

Net reaction $\qquad O + O_3 \rightarrow O_2 + O_2$ $\qquad (32)$

The M in reactions (28) and (30) is a third body that can be any other gas. But $O_3$ thermal decomposition may reverse the outcome of these photochemical reactions. Water vapor ($\lambda < 212$ nm), $CO_2$ ($\lambda < 227.5$ nm), and $SO_2$ ($\lambda < 210$ nm), and their thermal dissociation products (OH, CO, SO) may also be photolyzed by the enhanced UV flux of the early Sun. Qualitatively the situation is similar to that for the hot gas giant exoplanets called "roasters" such as HD 209458 and HD 189733b. These planets are very close to their primary stars (e.g., HD 209458b is 0.05 AU from its primary and receives 10,000 times the solar flux at Jupiter) where UV photons are being dumped into their hot tropospheres. There is a critical atmospheric level for each species where the thermochemical ($t_{thermo}$) and photochemical ($t_{photo}$) lifetimes are equal, below which $t_{thermo} < t_{photo}$ in the thermochemical zone and above which $t_{thermo} > t_{photo}$ in the photochemical zone (see section 5 of Visscher et al. 2006, Moses et al. 2011). The difference between Earth's silicate atmosphere and the "roaster" planets is that Earth is $O_2$-rich and oxidizing while the roasters are $H_2$-rich and reducing. A combined photochemical – thermochemical model is necessary to delineate the photochemically and thermochemically active zones of Earth's silicate vapor atmosphere. Again, in principle spectroscopic observations of hot rocky exoplanets can indirectly constrain chemistry on the hot, early Earth – something which was not even dreamed of a few years ago.

**Steam Atmosphere**

Arrhenius et al. (1974) were probably the first to propose that Earth had a steam atmosphere early in its history. They calculated that heating during accretion of the Earth should release water from the accreting material and lead to formation of a $H_2O$-bearing atmosphere and oceans after cooling. Radiative cooling calculations indicate cooling took

~ 2.5 Ma (Sleep et al. 2001). Shortly after the work of Arrhenius et al. (1974), Ahrens and colleagues experimentally studied impact degassing of the Murchison CM2 carbonaceous chondrite, carbonates, and hydrous minerals (e.g., Lange and Ahrens 1982a,b, 1986, Tyburczy and Ahrens 1985, 1987, 1988, Tyburczy et al. 1986a,b). Due in large part to their work, it is generally accepted that conversion of at least some of the kinetic energy of accreted planetesimals into heat led to impact degassing of volatiles during the accretion of the Earth. Water and $CO_2$ were the major species released by impact degassing of the materials Ahrens and colleagues studied, thus impact-generated atmospheres have historically been called steam atmospheres (e.g. Abe and Matsui, 1985; Lange and Ahrens, 1982a). Models of Earth's steam atmosphere and magma ocean were developed by Abe and Matsui in a series of pioneering papers (Abe and Matsui 1985, 1987, Matsui and Abe 1986). More recent work on the terrestrial magma ocean is described by Abe (1997), Abe et al. (2000), and Elkins-Tanton (2012). More recent work on the steam atmosphere is discussed by Zahnle et al. (1988, 2007, 2010) and Sleep et al. (2001).

Figure 5 shows the chemical equilibrium composition of the $H_2O$, C, and S gases at a constant total pressure of 1000 bars in Earth's steam atmosphere. This pressure corresponds to the complete BSE inventories of water, C, and S. As mentioned earlier, neither the solubility of water, $CO_2$, and S in molten peridotite nor the temperature-dependence of their solubilities in peridotite and silicate melts are well known. But if they have retrograde solubility with constant temperature coefficients of ~ 0.1 wt. % per 100 K, virtually no volatiles would be in super-liquidus temperature melts.

Figure 5 is plausibly an upper limit to the pressures of these gases in Earth's steam atmosphere. Zahnle et al. (1988) modeled steam atmospheres with 100-300 bars water, but did not consider $CO_2$ or sulfur. They correctly noted that "a fully self-consistent treatment would need to consider the partitioning of carbon between $CO_2$, CO, $CH_4$, and graphite and the solubilities of these species in the melt". Pawley et al (1992) found CO was insoluble in MORB basaltic melt at 1200 C and 500-1500 bar pressures but we are unaware of $CH_4$ solubility data for silicate melts. Again more experimental data on volatile solubilities in super-liquidus silicate melts appear to be needed.

Figure 5 illustrates several interesting points. Water vapor remains the major gas in the steam atmosphere until ~ 5700 K where OH becomes more abundant. At 6000 K OH and H have almost identical abundances. At temperatures below ~ 3600 K, $H_2O$, $CO_2$, and $SO_2$ are the three most abundant gases. Above this point OH becomes more important and displaces $SO_2$, $CO_2$, and finally $H_2O$ as a major gas. Dioxygen and $H_2$ are produced by thermal dissociation of water, but the $fO_2$ is only ~ 0.1 log unit larger than in a pure steam atmosphere because $O_2$ produced by thermal dissociation of water inhibits $O_2$ production by thermal dissociation of $CO_2$ and $SO_2$ to some extent. The production of H and $H_2$ reduces some sulfur to HS and $H_2S$, but both remain minor species. The abundances of $CH_x$ gases (e.g., CH, $CH_2$, $CH_3$, $CH_4$) are even smaller and are not shown on the graph.

The steam atmosphere, like the silicate atmosphere, is significantly more oxidizing (i.e., has a larger oxygen fugacity $fO_2$) than the $H_2$-rich solar nebula and as a result, easily oxidized elements (e.g., Cr, Mo, W as gaseous oxides and $H_2CrO_4$, $H_2MoO_4$, $H_2WO_4$ gases, Fe as $Fe(OH)_2$ vapor, B as $H_3BO_3$ (g), Si as $H_4SiO_4$ (g), V as gaseous oxides) may partition between the magma ocean and steam atmosphere. All of these elements occur in sublimate minerals that condensed out of terrestrial (steam-rich) volcanic gases at volcanic or fumarolic vents (see Table 1 in Brackett et al. 1995).

Sleep et al (2001) used PVT data on the NaCl – water system (Bischoff 1991) and an analogy with chemistry at modern mid-ocean ridges to discuss chemistry and water – rock interactions in the steam atmosphere. Depending on the pressure, the PVT data indicate either halite (at 754 K, 300 bar) or a NaCl-rich brine forms. Sleep et al (2001) also note that Cl-bearing amphiboles form at temperatures below ~ 1023 K in modern mid-ocean ridge axes. Schaefer et al. (2012) found that halite condenses at 1550 K, 100 bar in their modeling of the steam atmosphere produced by heating the bulk silicate Earth. Schaefer et al (2012) did not have thermodynamic data for endmember Cl-bearing amphiboles or micas for their calculations but Cl may substitute into the F- and OH-bearing micas and amphiboles, which do form. For example, at 100 bar total pressure we find that fluorphlogopite is stable below 1350 K and pargasite is stable below 1000 K in the steam atmosphere produced by heating the BSE. Thus, there is qualitative agreement between the steam atmosphere models of Schaefer et al. (2012) and the mid-ocean ridge analogy of Sleep et al (2001).

As noted above, the steam atmosphere lasted for ~ 2.5 Ma. After most of the steam rained out to form the oceans, Earth's atmosphere was plausibly dominated by "traditional" volatiles. Using Figure 5 as a guide, the major gases may have been mainly $CO_2$ with smaller amounts of $SO_2$, $H_2$, $H_2S$, and $H_2O$. The water vapor partial pressure was limited by its saturation vapor pressure and may have been ~ 50% of the saturated value as in Earth's troposphere today. Molecular $N_2$ was probably also abundant because $NH_3$ is so thermally unstable that it could not exist in an early steam atmosphere (see Table 5). Sleep and Zahnle (2001) showed that most of the atmospheric $CO_2$ (~ 250 bar in their work) could be lost by carbonization and subduction of rocks such as basalts or ultramafics. The timescale for loss of a $CO_2$-rich atmosphere is 10 – 100 Ma and depends on the details of their models.

However, it is also possible that the "traditional" volatiles left after collapse of the steam atmosphere were more reducing. Schaefer and Fegley (2010) showed that the "steam" atmosphere produced on either an ordinary chondritic (H, L, LL) or enstatite chondritic (EH, EL) early Earth was $H_2O$-poor and either $H_2$-rich (H, L, LL, EH chondritic material) or CO-rich (EL chondritic material). As noted earlier, several geochemical models predict the Earth formed from large amounts of ordinary and enstatite chondritic-like material, e.g., up to 70% enstatite chondrite-like material and up to 21% H-chondrite like material. The proto-Earth (and also the impactor) may have been more reducing than the present day Earth, and thus the resulting "steam" atmosphere may have been more reducing. If this were the case, the "traditional" volatiles left after collapse of the steam atmosphere may have been more reducing. For example, Table 5, based on Schaefer and Fegley (2010), shows that the gases remaining after collapse of the steam atmospheres on an ordinary or enstatite chondritic-like early Earth may have been $H_2$, CO, $CO_2$, and trace water vapor. We return to this point when we discuss outgassing on the early Earth.

**Impact degassing of the late veneer**

The abundances of highly siderophile elements (HSE, which are Os, Ir, Pt, Ru, Rh, Pd, Re, and Au) in the terrestrial mantle are larger than would be expected from equilibrium partitioning between mantle silicates and core-forming metal. Geochemists believe the HSE elements were added to the Earth as part of a late veneer of $\leq 1\%$ of Earth's mass at the end of its accretion (Palme and O'Neill 2003) after the Moon-forming impact. Traditionally the late veneer was assumed to be like CI carbonaceous chondrites (Chou 1978). However, recent measurements of the $^{187}Os/^{188}Os$ ratio in mantle xenoliths give a value for this ratio that is above that for carbonaceous chondrites but within the range of ordinary and enstatite chondrites (Palme and O'Neill 2003).

Schaefer and Fegley (2010) modeled chemistry of volatiles released during impact degassing of carbonaceous, enstatite, and ordinary chondritic material as a function of pressure and temperature. They found that degassing of CI and CM carbonaceous chondritic material produced $H_2O$-rich steam atmospheres in agreement with the results of impact experiments on the Murchison CM2 chondrite. However, degassing of other types of chondritic material produced atmospheres dominated by other gases. Degassing of ordinary (H, L, LL) and high iron enstatite (EH) chondritic material gave $H_2$-rich atmospheres with CO and $H_2O$ being the second and third most abundant gases at high temperatures, and increasing amounts of $CH_4$ at lower temperatures. Degassing of low iron enstatite (EL) chondritic material gave a CO-rich atmosphere with $H_2$, $CO_2$, and $H_2O$ being the next most abundant gases. Finally degassing of CV carbonaceous chondritic material gave a $CO_2$-rich atmosphere with $H_2O$ being the second most abundant gas. Their results at 1500 K, 100 bar total pressure are listed in Table 5, where the major gas from degassing of each type of chondritic material is in boldface.

Schaefer and Fegley (2010) did calculations over wide P, T ranges and showed that the results in Table 5 are generally valid at other pressures and temperatures. Water vapor, $CO_2$, and $N_2$ remain the major H-, C-, and N-bearing gases in atmospheres generated by degassing of CI and CM carbonaceous chondritic material, but $CH_4$ becomes the major C-bearing gas at low temperatures in atmospheres generated by degassing CV carbonaceous chondritic material. Methane also becomes a significant gas at low temperatures for degassing of CI chondritic material (see Fig. 2 of Zahnle et al. 2010). This surprising result is confirmed by the generally similar results of Hashimoto et al. (2007). Figures 3 and 5 in Schaefer and Fegley (2010) show a more complicated situation for atmospheres generated by impact degassing of ordinary (H, L,, LL) and enstatite (EH, EL) chondritic material. Either CO or $CH_4$ are the major C-bearing and either $H_2$ or $CH_4$ are the major H-bearing gases as P and T varies. Dinitrogen almost always is the major N-bearing gas because $NH_3$ is only stable in a small P-T range at low temperatures for degassing of H chondritic material. Holloway (1988) computed chemical equilibrium abundances for degassing of unequilibrated ordinary chondritic material and obtained very similar results at 50 bars pressure and 400 – 1200 $^o$C.

**Outgassing on the early Earth**

We now consider the speciation of carbon and nitrogen in volcanic gases on the early Earth. A good starting point is the chemistry of modern day volcanic gases (Symonds et al. 1994). Although their chemistry is spatially and temporally variable, some generalizations are possible. The three major gases and their mean concentration in 136 volcanic gas analyses for convergent plate, divergent plate, and hot spot volcanoes (Symonds et al. 1994) are $H_2O$ (80.4%), $CO_2$ (9.8%), and $SO_2$ (7.4%). Significant minor gases are $H_2$, CO, and $H_2S$, and common trace gases are OCS, $S_2$, HCl, and HF.

Present day volcanic gases are oxidized enough and hot enough that they emit neither $CH_4$ nor $NH_3$ (e.g., see analyses on pp. 14-18 of Symonds et al. 1994). When they do occur in modern day volcanic gases, $CH_4$ and $NH_3$ are probably due to deep hydrothermal or sedimentary gases that mix with the magmatic gas without equilibrating with it (see pp. 9-10 of Symonds et al. 1994).

Volcanic outgassing of $CH_4$ on the early Earth can be constrained via the reaction

$$CH_4 + 2O_2 = 2H_2O \text{ (steam)} + CO_2 \qquad (33)$$

This involves the implicit assumption that the major volatiles are the same as today, which is supported by Delano (2001) who showed that the mantle oxidation state has been approximately constant ($\pm$ 0.5 log unit $fO_2$) over the past 3,600 Ma. Using thermodynamic data from Table 10-10 of Fegley (2013), the standard Gibbs energy change ($\Delta G^o$ J mol$^{-1}$) of this reaction from 298 – 2500 K is

$$\Delta G^o = -798,305 + 7.4444T \log T - 24.6166T \qquad (34)$$

The corresponding equilibrium constant is given by

$$K_P = \exp\left(-\frac{\Delta G^o}{RT}\right) = \frac{P_{CO_2} P_{H_2O}^2}{P_{CH_4} P_{O_2}^2} \qquad (35)$$

Rearranging this expression we see that any given temperature, the oxygen partial pressure (fugacity) corresponding to equal partial pressures of $CH_4$ and $CO_2$ is given by the equation

$$f_{O_2} = \left[\left(\frac{X_{CO_2}}{X_{CH_4}}\right)\frac{X_{H_2O}^2 P_T^2}{K_P}\right]^{1/2} = \left[\frac{X_{H_2O}^2 P_T^2}{K_P}\right]^{1/2} \qquad (36)$$

The $X_i$ terms are mole fractions, $P_T$ is the total pressure of the volcanic gas, and $K_P$ is the equilibrium constant. Making the approximation that the $H_2O$ mole fraction is unity and that the total pressure is one bar we get a simple expression for the oxygen fugacity at which $CH_4$ and $CO_2$ are equally important in a volcanic gas

$$f_{O_2} = \left[ \frac{1}{K_P} \right]^{1/2} \qquad (37)$$

At 1400 K, a typical vent temperature at Kilauea, the oxygen fugacity is $10^{-14.92}$ bars for $CH_4$ and $CO_2$ to be equivalent. This is ~ 10 times lower than the $fO_2$ for the quartz – fayalite – iron (QFI) buffer, or the $fO_2$ for ordinary chondritic material at the same temperature (e.g., see Figure 8 in Schaefer and Fegley 2007). Equation (36) shows the estimated $fO_2$ is proportional to the $H_2O$ mole fraction and the total pressure. Exact values of the steam content and total pressure of primordial volcanic gases are not known, but are not essential to conclude that volcanic outgassing of $CH_4$ was probably insignificant on the early Earth. A similar constraint can be applied to $NH_3$ and it shows that volcanic outgassing of $NH_3$ also was insignificant. These conclusions are in accord with previous results (e.g., Abelson 1966, Delano 2001, Holland 1962, Rubey 1955).

However, $CH_4$ and $NH_3$ can be produced by outgassing at lower temperatures, even if the mantle redox state is the same as now. Schaefer et al. (2012) calculated chemical equilibrium abundances of gases outgassed by the bulk silicate Earth (BSE) over a wide P, T range as part of modeling the chemical composition of atmospheres expected on hot, rocky exoplanets. Figure 6 shows their results for the calculated oxygen fugacity of the BSE as a function of temperature at a constant total pressure of one bar. In general, the computed $fO_2$ of the heated BSE is about the same as the quartz – fayalite – magnetite (QFM) buffer. At magmatic temperatures the calculated $fO_2$ of the BSE is the same as the $fO_2$ of modern day volcanic gases (Symonds et al. 1994). At lower temperatures the $fO_2$ of the BSE falls below that of QFM, and is ~ 1 log unit lower at 550

K. The calculated results are consistent with the redox state of Earth's mantle which Frost and McCammon (2008) conclude is within ±2 log units of the QFM buffer.

Schaefer et al. (2012) found that $CH_4$ becomes the dominant C-bearing gas at low temperatures, e.g., 600 K at one bar total pressure. The exact temperature where the $CH_4/CO_2$ ratio is unity increases slightly with increasing total pressure and is ~ 700 K at 100 bars total pressure. Ammonia has qualitatively similar behavior, but at the same total pressure the $NH_3/N_2$ ratio reaches unity at a lower temperature than the $CH_4/CO_2$ ratio.

Frost and McCammon (2008) and Frost et al. (2008) note that during core formation the BSE was in equilibrium with Fe-Ni rich alloy and would have had an oxidation state ~ 4.5 log units below QFM at a given temperature, i.e. at the iron – wüstite buffer. The more reduced mantle would move the $CH_4 – CO_2$ crossover to higher temperatures at a given total pressure.

A significantly more reduced Earth would outgas a reduced atmosphere. At the suggestion of his advisor J. S. Lewis, Bukvic (1979) did chemical equilibrium calculations for gas-solid equilibria in the upper layers of an Earth-like planet that he modeled as H chondritic material or a mixture of 90% H and 10% CI chondritic material. He found outgassed atmospheres composed of $CH_4 + H_2$ in all cases. Schaefer and Fegley (2005, 2007) computed outgassing of average H chondritic material along the same thermal profile used by Bukvic (1979) and found a reduced atmosphere dominated by $CH_4$ and $H_2$ (see Figure 7). Similar results were obtained by Saxena and Fei (1988) who modeled outgassing of a carbonaceous chondritic planet. Schaefer and Fegley (2005) also studied outgassing from H chondritic material depleted in Fe metal and FeS (see Figure 8). A comparison of Figures 7 and 8 shows that a highly reduced atmosphere is produced in both cases, even after Fe metal and FeS have been removed, e.g., by core formation on the early Earth.

**Summary of Key Questions**

Some of the key questions about the chemistry of Earth's early atmosphere that remain unresolved are connected to the silicate and steam atmospheres on the early Earth. What is the temperature-dependent saturation vapor pressure and speciation in the bulk silicate Earth magma? We used the MAGMA code to discuss these issues, but Knudsen effusion mass spectrometric measurements of the saturated vapor pressure of molten peridotite and other molten silicates are needed to put models on a firm footing. The key conclusion from the MAGMA code that Si, $O_2$, and O are the major gases in the saturated vapor over the BSE magma is not at issue. But the temperature-dependent total vapor pressure and partial pressures of minor gases are less certain.

The temperature-dependent solubility of $H_2O$, $CO_2$, CO, $CH_4$, and sulfur in molten peridotite and other silicate melts at super-liquidus temperatures is another unresolved question. This will be difficult to measure but is important for modeling the silicate vapor and steam atmospheres on the early Earth and on rocky exoplanets.

The oxidation states of the silicate, steam, and "traditional" volatile atmospheres on the early Earth are another unanswered question. This can be addressed by a combination of thermochemical and photochemical models of the outgassing of different types of chondritic and achondritic material. The redox state of the proto-Earth during its accretion is an important factor determining the redox state of the outgassed atmospheres.

We close by returning to a point mentioned in the introduction; namely how spectroscopic observations of rocky exoplanets may provide indirect constraints on ideas about the Earth's early atmosphere. As noted earlier, as of the time of writing (Sept. 2012), there are three rocky exoplanets with sufficiently high surface temperatures to support silicate vapor atmospheres. These planets are CoRoT-7b (T ~ 2474 K, $1.58R_E$, $7.42M_E$, $\rho$ ~ 10.4 g cm$^{-3}$), Kepler-10b (T ~ 3038 K, $1.42R_E$, $4.54M_E$, $\rho$ ~ 8.7 g cm$^{-3}$), and 55 Cnc e (T ~ 2800 K, $2.0$-$2.1R_E$, $8.0$-$8.6M_E$, $\rho$ ~ 5.0-5.9 g cm$^{-3}$), where the sub-stellar temperatures are listed (see earlier discussion). The key gases to observe on these planets are SiO and $H_2O$. The presence of SiO gas proves the existence of a silicate vapor atmosphere and the presence (or absence) of $H_2O$ shows whether or not the planet is volatile depleted. Carter et al. (2012) recently reported the discovery of the rocky exoplanets Kepler-36b. This radius, mass, and density of Kepler-36b are ~ $1.49R_E$, ~ $4.45M_E$, and ~ 7.5 g cm$^{-3}$. It has a calculated surface temperature of ~ 980 K, which is hot enough to support a steam atmosphere. The key observation in this case (or of any other similar rocky exoplanets) is whether or not abundant $H_2O$ and $CO_2$ or $H_2$ and CO are present because these pairs of gases provide information on the oxidation state of the planet. In the ideal case a series of rocky exoplanets with surface temperatures ranging from ~ 300 – 2500 K could be observed spectroscopically. At present this lies in the future. But, the future may show that the answers to understanding Earth's early atmosphere lie not only in the rock record but also in astronomical observations of worlds beyond our solar system.

**Acknowledgments**

This work was supported by NASA cooperative agreement NNX09AG69A with the NASA Ames Research Center and by grants from the NSF Astronomy Program. We dedicate this chapter to the memory of our friend Al Cameron, who originally got B.F.

interested in the chemical consequences of the Moon-forming impact. We thank David Fike, Mark Marley, and Fred Moynier for their comments, and Jim Tyburczy and Kevin Zahnle for their reviews of the manuscript, and James Farquhar for his extraordinary patience while waiting for this chapter.

Table 1. Volatile inventories and depletion factors on the Earth

| Volatile | Solar Abundance[1] | µg/g in BSE[2] | Inventory (kg) | Depletion factor | Notes[3] |
|---|---|---|---|---|---|
| H (water) | $1.27 \times 10^7$ | 1072 | $4.32 \times 10^{21}$ | $6.2 \times 10^{-4}$ | solar $A_{water} = A_O - A_{Mg} - 2A_{Si}$, adjusted for O in rock |
| | | | | | BSE water calculated from 120 µg/g H in BSE |
| C | $7.19 \times 10^6$ | 100 | $4.03 \times 10^{20}$ | $1.5 \times 10^{-4}$ | C in BSE is 46-250 µg/g, see Table 6.9 of LF98 |
| N | $2.12 \times 10^6$ | 2 | $8.06 \times 10^{18}$ | $1.9 \times 10^{-5}$ | Atmosphere ~ 50% of total N in BSE |
| F | 804 | 25 | $1.01 \times 10^{20}$ | 0.22 | F in BSE is 19-28 µg/g, see Table 6.9 of LF98 |
| Ne | $3.29 \times 10^6$ | $1.6 \times 10^{-5}$ | $6.50 \times 10^{13}$ | $7.6 \times 10^{-11}$ | Taking atmospheric Ne as the total inventory |
| S | $4.21 \times 10^5$ | 124 | $5.00 \times 10^{20}$ | $1.2 \times 10^{-3}$ | S in BSE is 13-1000 µg/g, see Table 6.9 of LF98 |
| Cl | 5170 | 30 | $1.21 \times 10^{20}$ | $2.2 \times 10^{-2}$ | Cl in BSE is 8-44 µg/g, see Table 6.9 of LF98 |
| $^{36+38}$Ar | $9.27 \times 10^4$ | $6.0 \times 10^{-6}$ | $2.40 \times 10^{13}$ | $2.3 \times 10^{-10}$ | Taking atmospheric Ar as total inventory of $^{36+38}$Ar |
| Kr | 55.8 | $4.2 \times 10^{-6}$ | $1.69 \times 10^{13}$ | $1.2 \times 10^{-7}$ | Taking atmospheric Kr as the total inventory |
| Xe | 5.46 | $5.0 \times 10^{-7}$ | $2.03 \times 10^{12}$ | $9.3 \times 10^{-8}$ | Taking atmospheric Xe as the total inventory |

[1] Solar abundance per $10^6$ Si atoms, Table 1.2 of Lodders, K. and Fegley, B., Jr. (2011). *Chemistry of the solar system*, Cambridge: RSC Publishing.

[2] Concentrations in the bulk silicate Earth (BSE) for H, C, N, F, and Cl are from Palme, H. and O'Neill, H. St. C (2003). Cosmochemical estimates of mantle composition. In *Treatise on geochemistry*, vol. 2. Sulfur is from Table 4.4 of Lodders and Fegley (2011)

[3] The range of published estimates for H, C, N, F, S, and Cl are in Table 6.9 of Lodders, K. and Fegley, B., Jr. (1998). *The planetary scientist's companion*, New York: Oxford University Press. The solar abundance of water is calculated from the oxygen abundance adjusted for the amount of oxygen in rock ($MgO + SiO_2$). Other values that are used in the calculations are the Si concentration in the BSE (21.22%), the mean molecular weight of Earth's atmosphere (28.97 g mol$^{-1}$), total atmospheric mass ($5.137 \times 10^{18}$ kg), mass of the BSE ($4.03 \times 10^{24}$ kg), and the concentrations of Ne, Ar, Kr, and Xe in dry air (18.18 ppmv, 9340 ppmv, 1.14 ppmv, and 87 ppbv). The Ar abundance in air is corrected for $^{40}$Ar, which is 99.6% of terrestrial Ar. Calculations compare terrestrial and solar abundances of $^{36}$Ar and $^{38}$Ar.

Table 2 Some volatile-bearing phases in chondrites and potential outgassed volatiles

| Name | Ideal Chemical Formula | Chondrite | Potential volatiles[b] |
|---|---|---|---|
| Apatite | $Ca_5(PO_4)_3(F,Cl,Br,OH)$ | many | HF, HCl, $Cl_2$, HBr, $Br_2$, $H_2O$, $H_2$, $O_2$ |
| Calcite | $CaCO_3$ | C, | CO, $CO_2$ |
| Cohenite | $(Fe,Ni)_3C$ | many | $CH_4$, CO, $CO_2$ |
| Dolomite | $CaMg(CO_3)_2$ | C | CO, $CO_2$ |
| Graphite | C | C | $CH_4$, CO, $CO_2$ |
| Gypsum | $CaSO_4 \cdot 2H_2O$ | C, OC | $SO_2$, $H_2S$, OCS, $S_x$ |
| Halite | NaCl | C, OC | HCl, $Cl_2$ |
| Insoluble organic matter | $C_{100}H_{72}N_3O_{22}S_{4.5}$ | C, UOC | $CH_4$, CO, $CO_2$, $H_2O$, $H_2$, $N_2$, $NH_3$, $S_x$, $H_2S$, OCS, $SO_2$ |
| Nierite | $Si_3N_4$ | E | $N_2$, $NH_3$ |
| Osbornite | TiN | E, CH | $N_2$, $NH_3$ |
| Sinoite | $Si_2N_2O$ | E | $N_2$, $NH_3$ |
| Serpentine | $(Mg,Fe)_3Si_2O_5(OH)_4$ | C | $H_2O$, $H_2$, $O_2$ |
| Sodalite | $Na_4Al_3Si_3O_{12}Cl$ | C | HCl, $Cl_2$ |
| Talc | $(Mg,Fe)_3Si_4O_{10}(OH)_2$ | C | $H_2O$, $H_2$, $O_2$ |
| Troilite | FeS | many | $S_x$, $H_2S$, OCS, $SO_2$ |

[a]The abbreviations denote the following types of chondrites: C = carbonaceous chondrites, CH = CH chondrites, E = enstatite (EH, EL) chondrites, OC = ordinary (H, L, LL) chondrites, UOC = unequilibrated ordinary chondrites, or many for phases found in many types of chondrites.

[b]The nature of the potential outgassed volatiles depends on several factors including the temperature, pressure, and oxygen fugacity during outgassing. Elemental fluorine does not form because it is too reactive. Hydrogen and oxygen are generated via equilibria of water vapor with Fe-bearing phases such as metal, magnetite, and FeO-bearing silicates.

Table 3 Thermal effects of large impacts on the early Earth

| $\Delta E$ (J) | Size of impactor | Thermal effects for Earth |
|---|---|---|
| $7 \times 10^{27}$ | $1.4 \times 10^{20}$ kg (~mass of asteroid 2 Pallas) | Boil oceans and heat to 2000 K |
| $5 \times 10^{28}$ | $1 \times 10^{21}$ kg (~mass of asteroid 1 Ceres) | Melt crust and heat to 2000 K |
| $2 \times 10^{29}$ | $4 \times 10^{21}$ kg (~5% mass of Earth's moon) | Vaporize crust and heat to 3200 K |
| $3 \times 10^{31}$ | $6.8 \times 10^{23}$ kg (~mass of Mars) | Vaporize silicate Earth and heat to 3690 K |

Table 4. Composition of the bulk silicate Earth

| Oxide | Wt % |
|---|---|
| $SiO_2$ | 45.86 |
| $MgO$ | 37.12 |
| $Al_2O_3$ | 4.55 |
| $TiO_2$ | 0.215 |
| $FeO$ | 8.18 |
| $CaO$ | 3.69 |
| $Na_2O$ | 0.353 |
| $K_2O$ | 0.031 |
| Total | 100.00 |

Table 5. Major Gas Compositions of Impact-generated Atmospheres from Chondritic Planetesimals at 1500 K and 100 bars

| Gas (vol %) | CI | CM | CV | H | L | LL | EH | EL |
|---|---|---|---|---|---|---|---|---|
| $H_2$ | 4.36 | 2.72 | 0.24 | **48.49** | **42.99** | **42.97** | **43.83** | 14.87 |
| $H_2O$ | **69.47** | **73.38** | 17.72 | 18.61 | 17.43 | 23.59 | 16.82 | 5.71 |
| $CH_4$ | $2\times10^{-7}$ | $2\times10^{-8}$ | $8\times10^{-11}$ | 0.74 | 0.66 | 0.39 | 0.71 | 0.17 |
| $CO_2$ | 19.39 | 18.66 | **70.54** | 3.98 | 5.08 | 5.51 | 4.66 | 9.91 |
| CO | 3.15 | 1.79 | 2.45 | 26.87 | 32.51 | 26.06 | 31.47 | **67.00** |
| $N_2$ | 0.82 | 0.57 | 0.01 | 0.37 | 0.33 | 0.29 | 1.31 | 1.85 |
| $NH_3$ | $5\times10^{-6}$ | $2\times10^{-6}$ | $8\times10^{-9}$ | 0.01 | 0.01 | $9\times10^{-5}$ | 0.02 | $5\times10^{-5}$ |
| $H_2S$ | 2.47 | 2.32 | 0.56 | 0.59 | 0.61 | 0.74 | 0.53 | 0.18 |
| $SO_2$ | 0.08 | 0.35 | 7.41 | $1\times10^{-8}$ | $1\times10^{-8}$ | $3\times10^{-8}$ | $1\times10^{-8}$ | $1\times10^{-8}$ |
| Other[a] | 0.25 | 0.17 | 1.02 | 0.33 | 0.35 | 0.41 | 0.64 | 0.29 |
| Total | 99.99 | 99.96 | 99.95 | 99.99 | 99.97 | 99.96 | 99.99 | 99.98 |

[a]Other includes gases of the rock-forming elements Cl, F, K, Na, P, and S.

Figure Captions

Figure 1. Depletion factors for inert gases and chemically reactive volatiles on Earth relative to their abundances in the solar nebula. See Table 1 and the text for details.

Figure 2. Average abundances of water, carbon, and nitrogen in ordinary (OC), enstatite (EC), carbonaceous (CC) chondrites, howardite, eucrite, and diogenite meteorites (HED) and the Earth.

Figure 3. Temperature-dependent vapor pressures of iron metal, forsterite, fayalite, silica, and bulk silicate Earth magma are compared to the total partial pressure of $H_2O$ + C, S gases corresponding to the BSE inventory of these volatiles.

Figure 4. Temperature-dependent chemical equilibrium composition of the saturated vapor in equilibrium with bulk silicate Earth magma.

Figure 5. Temperature-dependent chemical equilibrium composition of Earth's steam atmosphere at 1000 bar total pressure.

Figure 6. Temperature-dependent oxygen fugacities for modern day volcanic gases and the heated bulk silicate Earth (Schaefer et al. 2012) are compared to oxygen fugacities of several common buffers denoted by the following abbreviations: MH = magnetite – hematite, NNO = Ni – NiO, QFM = quartz – fayalite – magnetite, WM = wüstite – magnetite, IM = iron – magnetite, and IW = iron – wüstite.

Figure 7. Chemical equilibrium abundances of gases produced by heating average H chondritic material along a terrestrial geotherm.

Figure 8. Same as in Figure 7, but after removal of all Fe metal and FeS before doing the computations.

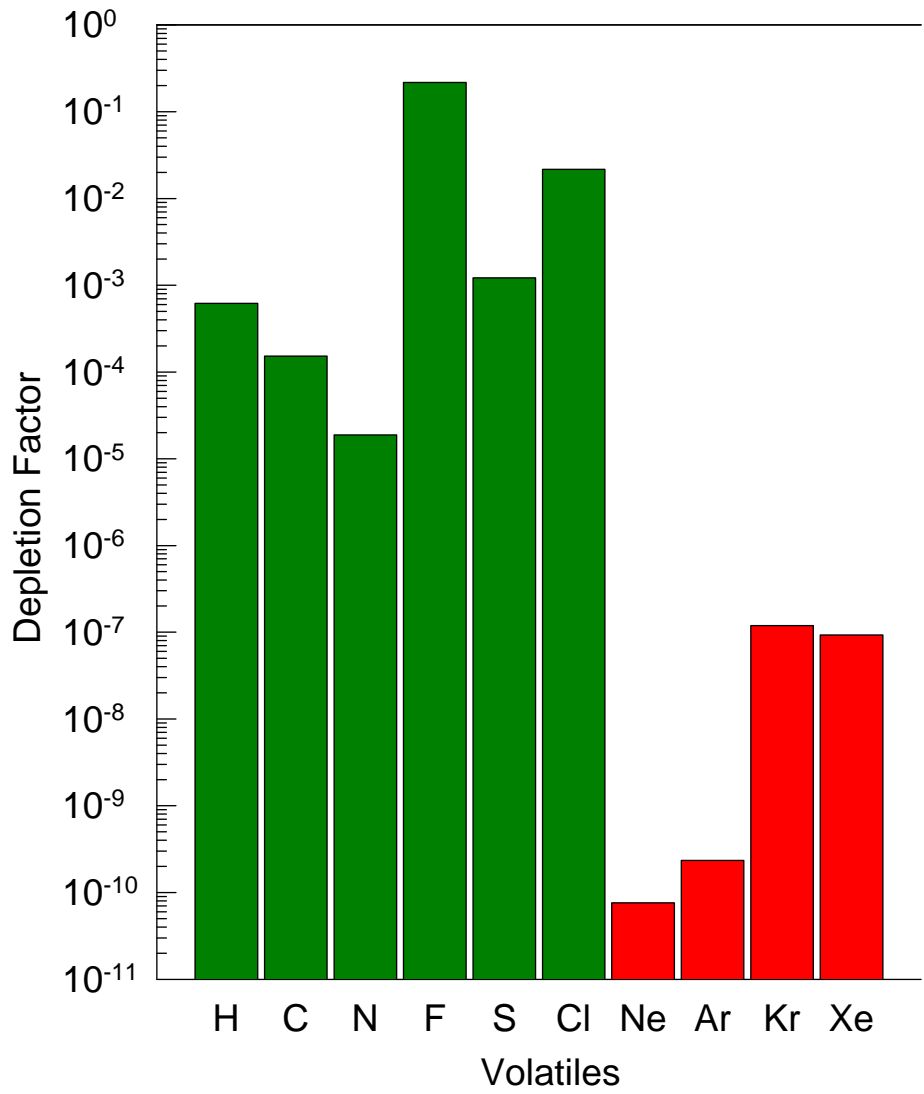

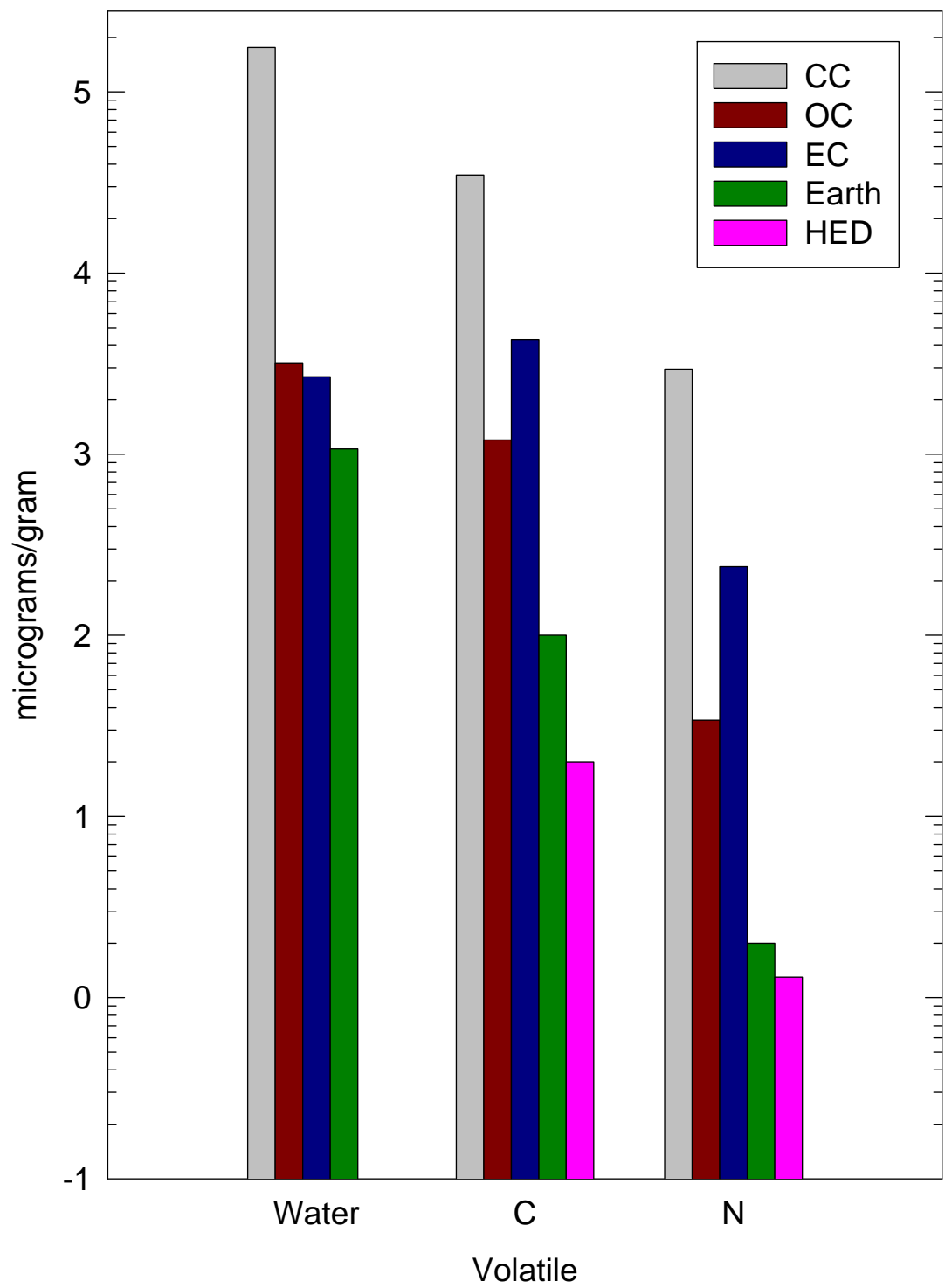

Figure 4.24

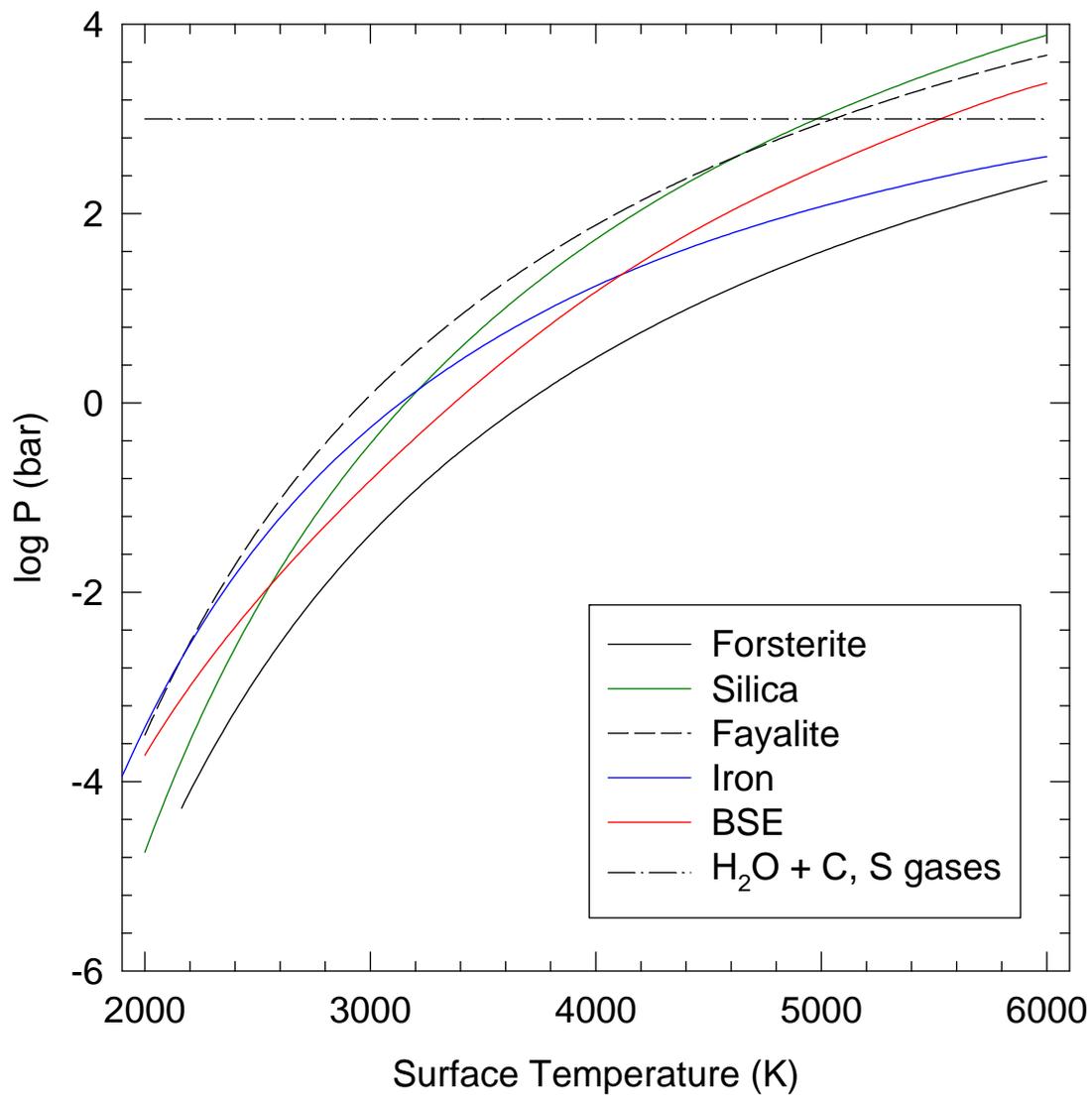

vaporPcv.spw

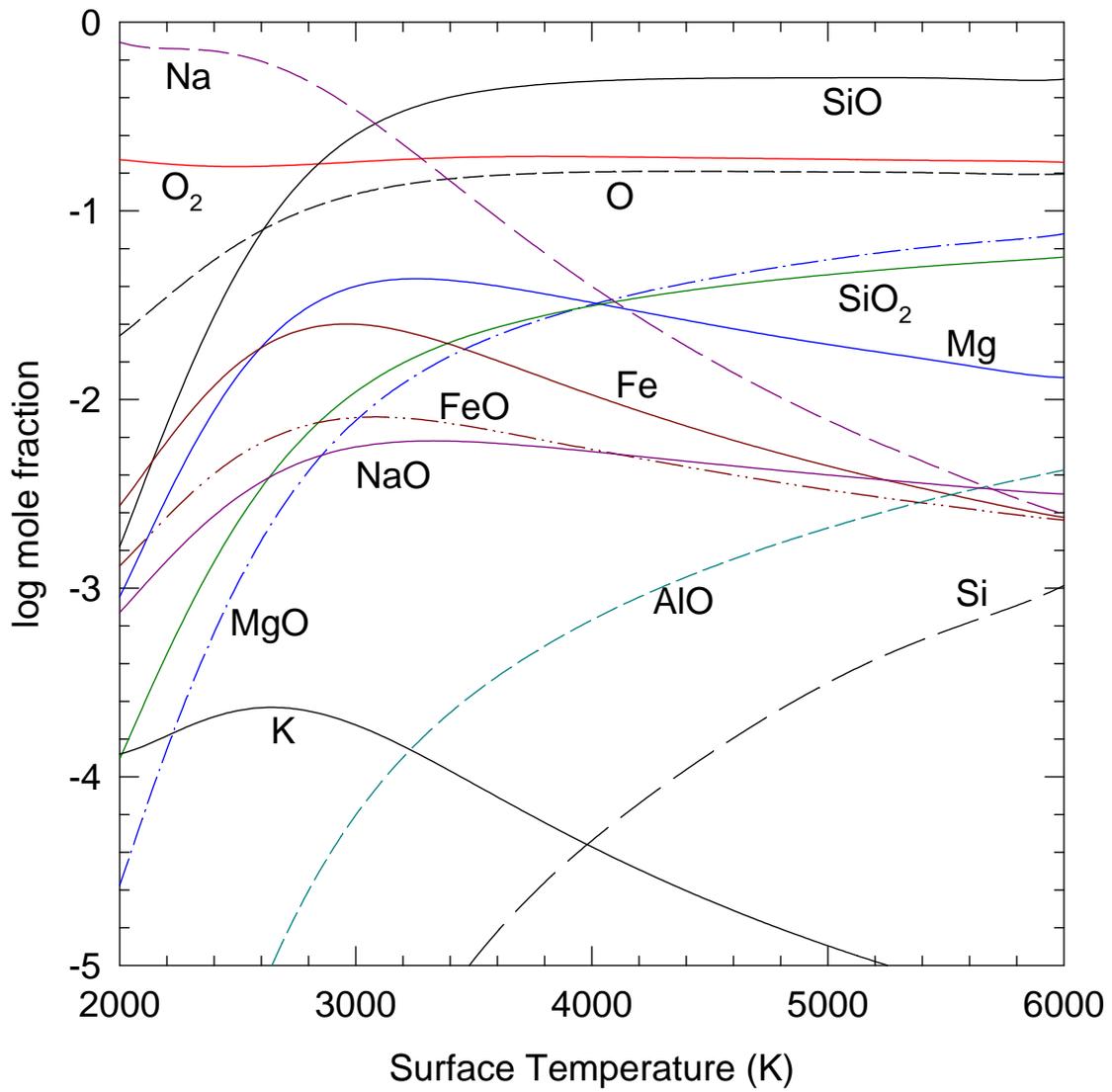

**Composition of saturated vapor
Bulk silicate Earth magma**

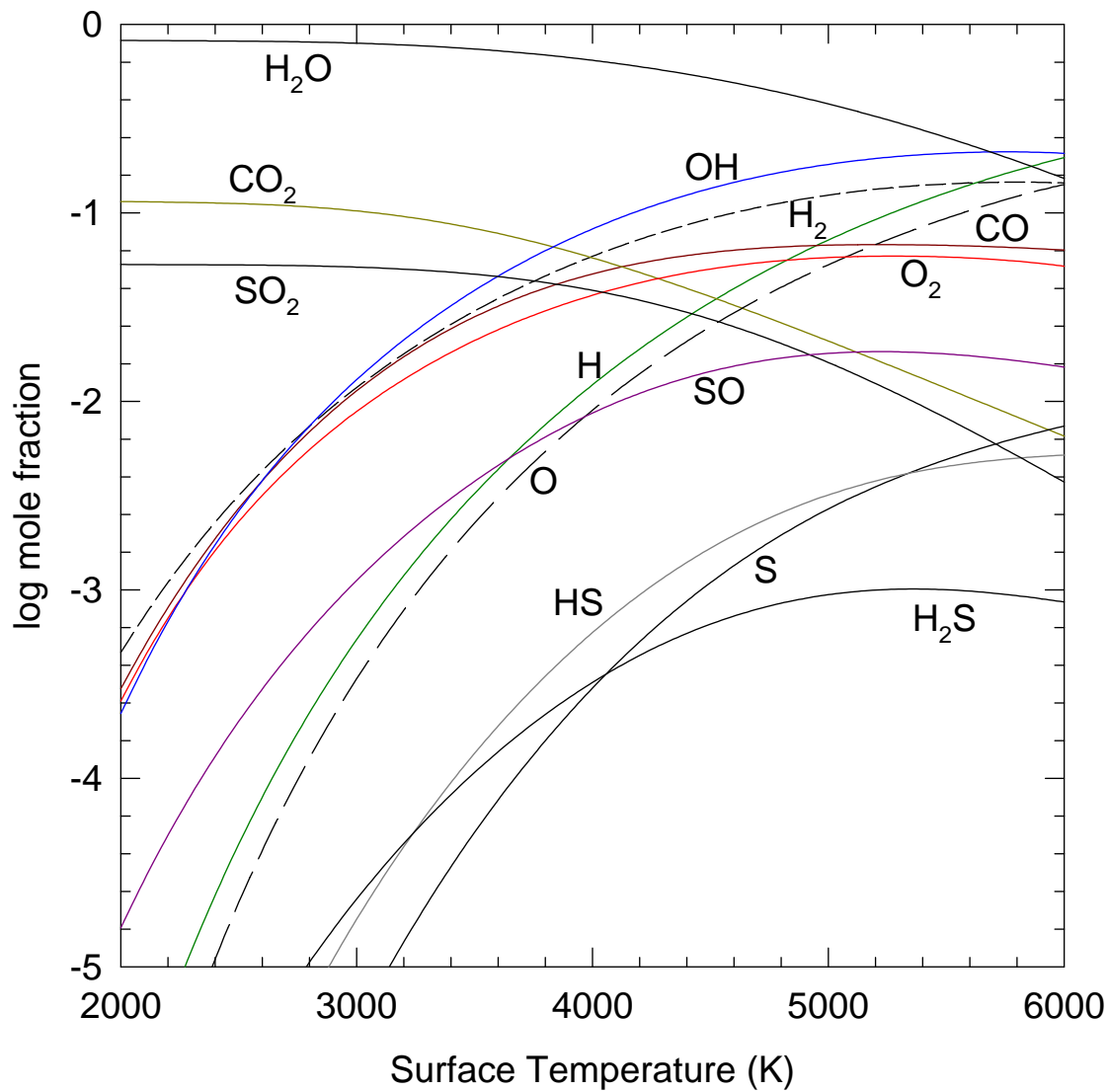

**Steam - CO$_2$ - SO$_2$ Atmosphere**
**P$_T$ = 1000 bar**

CHOSAtm.SPW

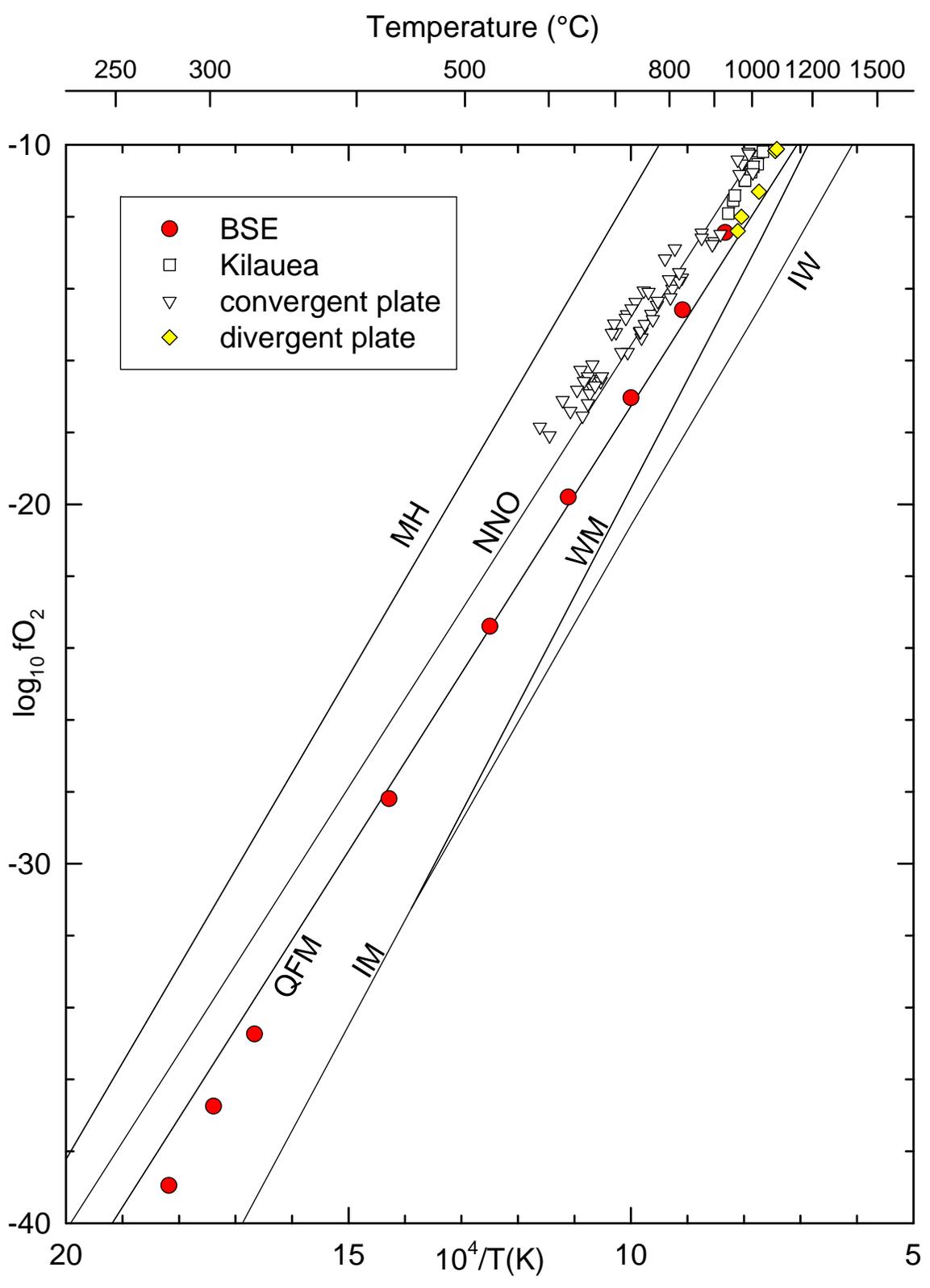

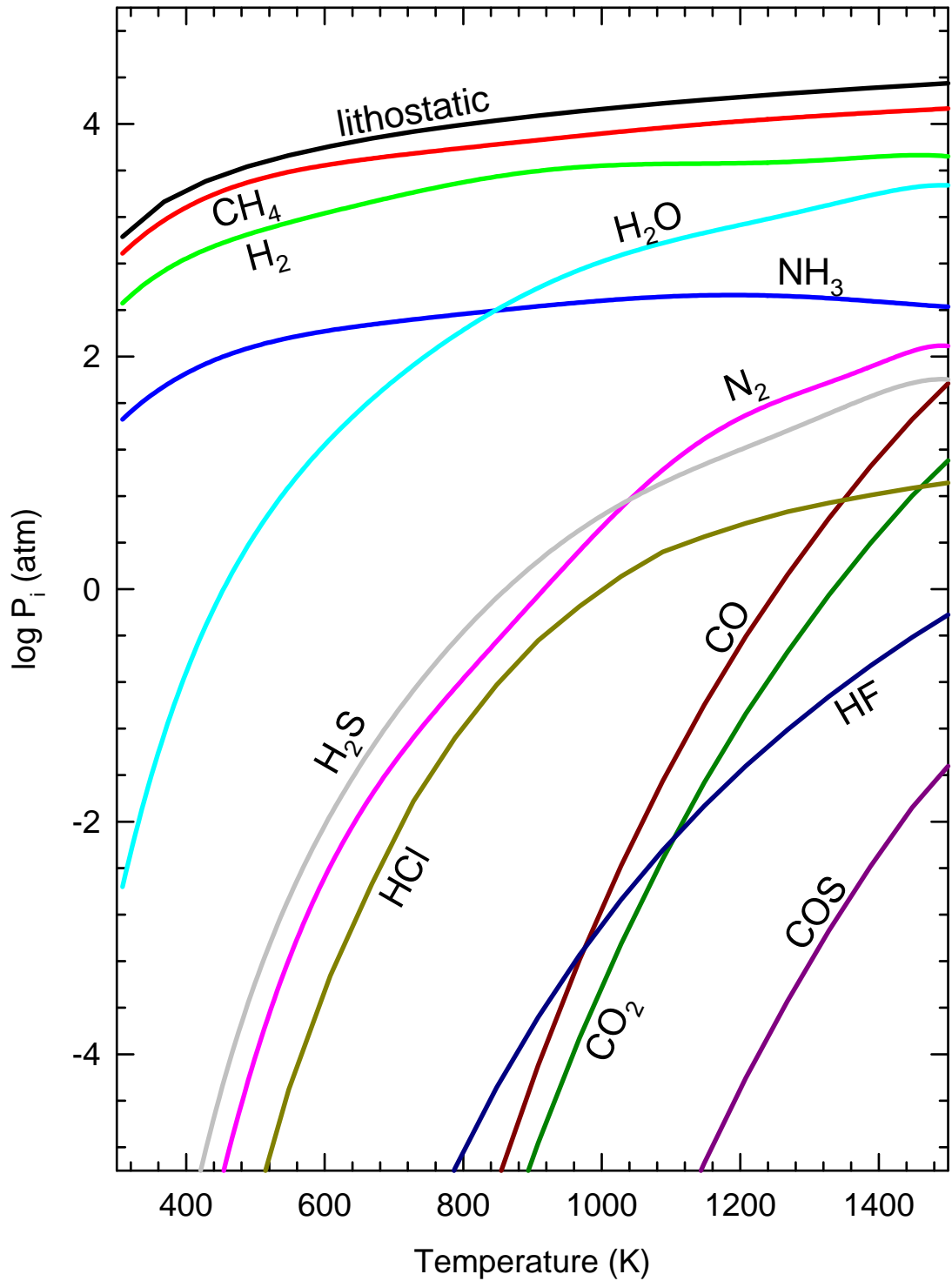

**Average H (falls) composition
from Jarosewich 1990**

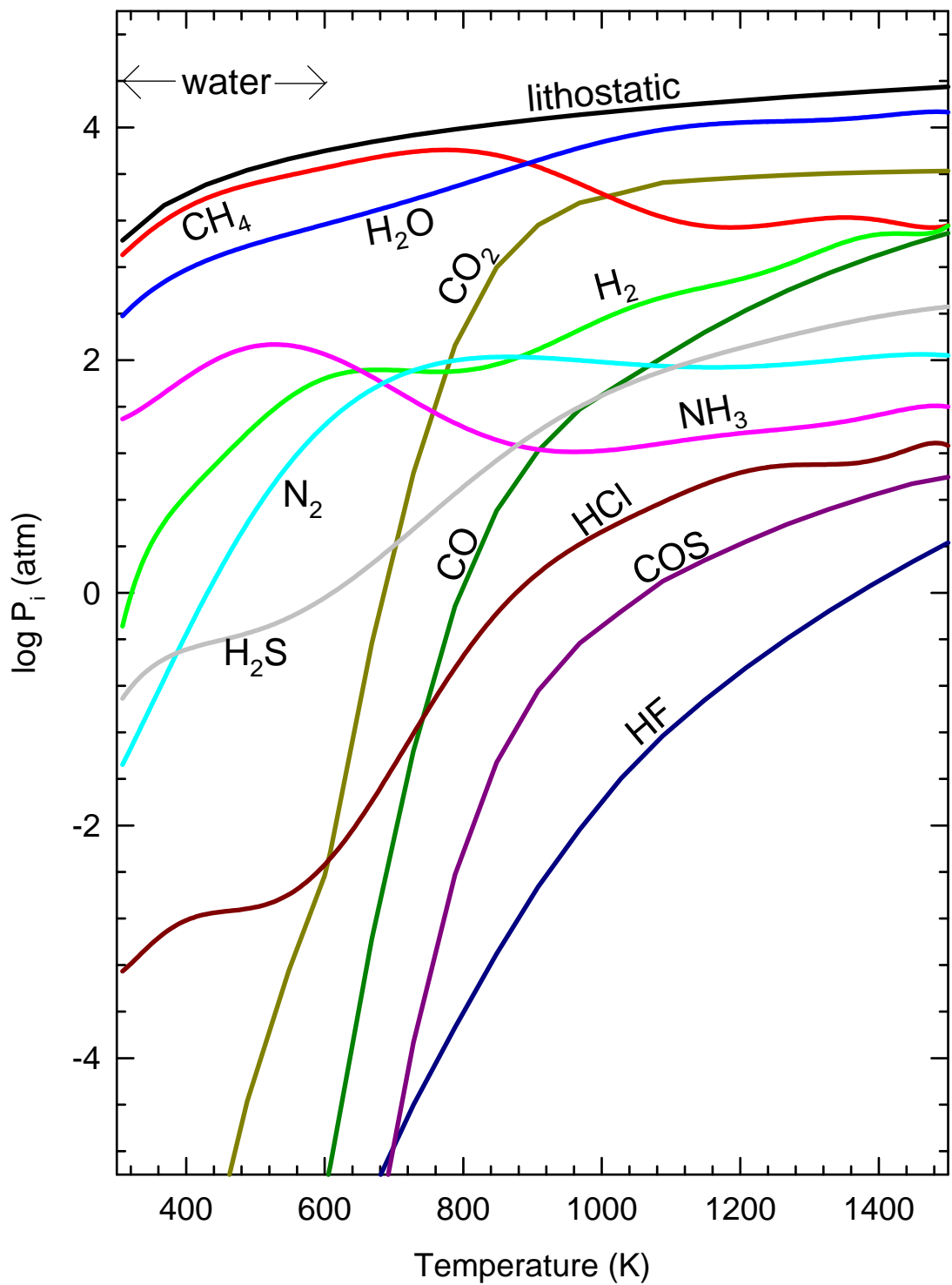

**Average H (falls) - no Fe metal or FeS (S = 10% total S)**
**from Jarosewich (1990)**

Figure axes: $\log P_i$ (atm) versus Temperature (K)

Labeled curves: lithostatic, $CH_4$, $H_2O$, $CO_2$, $H_2$, $N_2$, $NH_3$, $H_2S$, $CO$, $HCl$, $COS$, $HF$, water